\documentclass[conference]{IEEEtran}
\IEEEoverridecommandlockouts

\usepackage{cite}
\usepackage[inline]{enumitem}
\usepackage{algorithmic}
\usepackage{amsmath,amssymb,amsfonts}
\usepackage{multirow}
\usepackage[linesnumbered,ruled,vlined]{algorithm2e}
\usepackage{adjustbox}
\usepackage{xcolor}
\usepackage{booktabs}
\usepackage[hidelinks]{hyperref}

 \def\BibTeX{{\rm B\kern-.05em{\sc i\kern-.025em b}\kern-.08em
     T\kern-.1667em\lower.7ex\hbox{E}\kern-.125emX}}

\DeclareMathOperator*{\argmin}{arg\,min}

\newcommand{\method}[0]{FRAUDability\xspace}
\graphicspath{{./figures/}}

\begin{document}

\title{FRAUDability: Estimating Users' Susceptibility to Financial Fraud Using Adversarial Machine Learning}

\author{\IEEEauthorblockN{Chen Doytshman\IEEEauthorrefmark{1}, Satoru Momiyama\IEEEauthorrefmark{2}, Inderjeet Singh\IEEEauthorrefmark{2}, 
Yuval Elovici\IEEEauthorrefmark{1}, Asaf Shabtai\IEEEauthorrefmark{1}}
\IEEEauthorblockA{\IEEEauthorrefmark{1}Department of Software and Information Systems Engineering, \\ Ben-Gurion University of the Negev, Beer-Sheva}
\IEEEauthorblockA{\IEEEauthorrefmark{2}NEC Corporation}}


\maketitle

\begin{abstract}
In recent years, financial fraud detection systems have become very efficient at detecting fraud, which is a major threat faced by e-commerce platforms.
Such systems often include machine learning-based algorithms aimed at detecting and reporting fraudulent activity.
In this paper, we examine the application of adversarial learning based ranking techniques in the fraud detection domain and propose FRAUDability, a method for the estimation of a financial fraud detection system's performance for every user.
We are motivated by the assumption that "not all users are created equal" -- while some users are well protected by fraud detection algorithms, others tend to pose a challenge to such systems.
The proposed method produces scores, namely ``fraudability scores'', which are numerical estimations of a fraud detection system's ability to detect financial fraud for a specific user, given his/her unique activity in the financial system.
Our fraudability scores enable those tasked with defending users in a financial platform to focus their attention and resources on users with high fraudability scores to better protect them.
We validate our method using a real e-commerce platform's dataset and demonstrate the application of fraudability scores from the attacker's perspective, on the platform, and more specifically, on the fraud detection systems used by the e-commerce enterprise.
We show that the scores can also help attackers increase their financial profit by 54\%, by engaging solely with users with high fraudability scores, avoiding those users whose spending habits enable more accurate fraud detection.
\end{abstract}





\section{\label{sec:intro}Introduction}

\begin{figure}[ht]
\includegraphics[width=\linewidth]{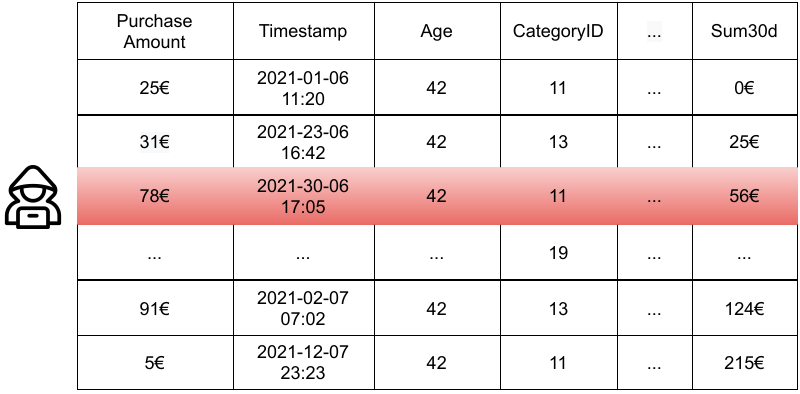}
\caption{After compromising a user's account, an attacker executes a transaction on behalf of the victim.}
\label{fig:fraud_in_sequence}
\end{figure}

Machine learning (ML) has become a first class citizen in the financial fraud detection domain, where ML based fraud detectors have obtained promising results~\cite{mercer1990fraud,22_malekian2013adaptive,jurgovsky2018sequence}.
These detectors are trained to detect transactions performed by an attacker (i.e., a fraudster) that has managed to compromise a user's account (e.g., by stealing the user's credentials), as illustrated in Figure~\ref{fig:fraud_in_sequence}.

Despite their high degree of accuracy, we argue that ML based fraud detectors do not perform equally well for all users, mainly because of the variability among users' financial behavior; for some users, these detectors are able to detect fraudulent transactions more accurately, while for others, the detectors are more prone to error.
In this study we introduce the notion of a \textit{fraudability score}, where users with a higher fraudability score are those who fraud detectors perform poorly on.
As a consequence, fraudulent activity performed against those users is difficult to detect, they are more attractive to attackers.
The ability to provide an estimation of a user's fraudability score can be used to distinguish more ``fraudable" and less ``fraudable" users.
Given this, we propose FRAUDability, an ML based method for the estimation of the ability of fraud detection systems to detect financial fraud for each user, namely the fraudability score.

Adversaries could use the fraudability score as a preliminary step before launching an attack on a compromised user account, assuring a higher success rate and a lower chance of detection. 
From the defender standpoint, service providers could use the proposed method to focus the attention of fraud detection systems on the users for whom the fraud detection system's performance is poorer; this would allow service providers to implement additional defenses, such as two-factor authentication and human expert analysis, or set lower thresholds for their fraud detection systems for these users specifically.

Previous research on the security and trustworthiness of ML based fraud detection systems mainly focused on impairing the performance of instance based detectors by crafting \textit{adversarial examples} -- inputs that are intentionally designed to cause the fraud detector to ignore the fraudulent transaction~\cite{DBLP:conf/raid/CarminatiSPZ20,DBLP:conf/www/GuoLAHHZZ19}.
However, these studies did not consider the difference in the performance (i.e., success rate) of the proposed adversarial ML attacks on different users.
In this paper, our goal is develop a method capable of estimating the ability of fraud detectors to detect financial fraud in users' accounts as a means of identifying users that are more fraudable, or in other words, more vulnerable to attacks.
We achieve this goal by simulating real fraud, including fraud generated by an adversarial ML attack.
Unlike prior studies, we focus mainly on sequence based fraud detection models, as this domain is relatively unexplored.

FRAUDability is a multi-stage method (an overview of proposed method is presented in Figure~\ref{fig:method}).
In the first stage, a subset of users is selected from all potential users. 
A defender may use a wide distribution of users, representing the entire user population, while an attacker may select all compromised user accounts~\cite{continella2017prometheus} (stage 1).
Next, the financial activity of the selected users is monitored and logged.
The logged transactions of the monitored users are used to create a surrogate dataset (stage 2).
We assume that the majority of the transactions in the surrogate dataset are benign.  
Then, the surrogate dataset is used to train a surrogate fraud detector (stage 3).
This surrogate detector is able to classify a given sequence of transactions as fraudulent or benign, similar to the fraud detectors used by service providers.
Next, a labeled dataset for training the fraudability score predictor is generated (stage 4).
In order to generate the labeled dataset, various fraudulent transactions are artificially injected into the transaction sequences of the users in the surrogate dataset (collected in stage 2).
We used various injection strategies, including random injection, as well as a novel sequence based adversarial learning attack that determines the location and attributes of the injected transaction. 
For each injected transaction, the surrogate fraud detector (trained in stage 3) is applied to evaluate whether the detector will detect the fraudulent transaction.
The amount of stolen money (i.e., financial profit gained from undetected fraudulent transaction) serves as the basis of the fraudability score (i.e., the label) of each user.
The fraudability score predictor model is trained using the fraudability scores of the users and various aggregated features that are extracted for each user (stage 5).
This model can predict the fraudability score of a new user based on the user's past activity.
Finally, the trained model is used to rank new users, assigning each user a score representing his/her susceptibility to financial fraud. 
This score is an estimation of the ability of the fraud detection system to detect financial fraud for that user (stage 6).

We evaluated our proposed method using a propriety dataset containing more than three million online transactions performed by more than 500,000 users on a real e-commerce platform.
Our experimental evaluation was \textit{conducted from an attacker's point of view}, i.e., the goal of the evaluation is to understand the extent to which the ability to predict users' fraudability scores can help an attacker perform fraudulent transactions without being detected.
We compared the attack success (evasion) rate when FRAUDability was used to select the users to attack (i.e., perform fraudulent transactions on their accounts) to the success rate of an attack when a random selection of users was applied.
The results show that when FRAUDability was used, the attacker was able to improve the success rate by 58\% for the random injection strategy and by 19\% for the proposed adversarial attack injection strategy.
In terms of profit, the improvement is even more significant, with a 166\% increase in the money stolen with the random injection strategy and a 30\% increase with the adversarial attack injection strategy.
The FRAUDability score predictor was able to predict the fraudabilty scores of new users with a low mean absolute error, even after trained on only one week worth of transactions.
We also show that our proposed sequence based adversarial attack injection strategy achieves high rates (92-98\%) of undetected fraudulent transactions.

The main contributions of this research can be summarized as follows:
\begin{itemize}
  \item Introducing the concept of a financial fraudability score, which is motivated by our evaluation results that show that fraud detectors do not perform equally on all users.
  \item An ML based method for the estimation of a fraud detection system's performance on different users.
  This method can be used to assign each user a ``fraudability score."
  \item An adversarial learning technique inspired by the fast gradient sign method (FGSM) and the basic iterative method (BIM) which are used in this work, to craft adversarial transactions aimed at fooling \textit{sequence-based} ML fraud detectors.
  
\end{itemize} 
\section{\label{sec:related}Related Work}

While the growth of e-commerce provides new business opportunities, as demonstrated during the COVID-19 pandemic, it has also become a playground for fraudsters who abuse the transparency of online payment methods and credit card transactions.
The rise in credit card fraud has had great impact on the finance sector: According to~\cite{17_thenilsonreport}, in 2017, global credit card fraud loss came to the staggering amount of \$27.85 billion.
By 2023, credit card fraud loss is projected to reach \$35.67 billion.

As a result, the finance sector has sought solutions, and ML has been examined extensively in this context~\cite{12_qi2018fintech}, particularly in the fraud detection domain~\cite{13_abdallah2016fraud,10_mozaffari2014systematic,15_randhawa2018credit}.
Most prior studies implemented instance based models, where each transaction is represented using a set of extracted features and the model classifies each transaction separately.
For example, Bhattacharyya et al.~\cite{14_bhattacharyya2011data} evaluated two data mining approaches, namely support vector machines (SVMs) and random forests, together with the well-known logistic regression, in an attempt to detect credit card fraud.
Their experiments showed that all of the techniques had the ability to model fraud in the considered data. 
Can et al.~\cite{DBLP:conf/kdd/LiuZASLFHT20} suggested a behavior-driven approach to detect fraud. 
They devised a tree-like structure referred to as a behavior tree to model the user's behavioral data. 
Tested on the Alibaba platform dataset, the proposed approach outperformed state-of-the-art methods.
We argue that the analysis of a sequence of transactions is important for the accurate detection of fraudulent transactions.
Therefore, unlike previous studies, our proposed experimental fraud detector is sequence-based, i.e., receives and classifies sequences of transactions instead of individual transactions.

In recent years ML models have been shown to be vulnerable to adversarial machine learning (AML) attacks; these attacks have become a major threat and are being developed at a rapid rate. 
Researches have demonstrated that with a minor change to the input data, one can deceive the ML algorithm so that it provides unexpected results~\cite{2_carlini2017towards,DBLP:journals/corr/GrossePM0M16}.
Other studies have demonstrated the capabilities of attacks employing adversarial examples in a wide range of domains like voice recognition, spam filtering, and autonomous cars~\cite{DBLP:journals/corr/abs-1801-01944,5_nelson2008exploiting, DBLP:journals/corr/abs-1903-05157}.

AML attacks have been explored in the domain of fraud detection as well.
Carminati et al.~\cite{DBLP:conf/raid/CarminatiSPZ20} studied the application of AML techniques in the banking fraud detection domain.
Assuming an attacker that has a dataset of old transactions and can execute transactions on behalf of the victim, the authors describe a process of crafting adversarial examples that can evade the target fraud detector.
Their approach is composed of a training phase, in which an oracle fraud detector is trained, and a runtime phase, in which evasive transactions are generated.
In their work, adversarial transactions are generated essentially by changing the mutable features of each transaction, e.g., \textit{amount} and \textit{timestamp}.
Evaluated with two proprietary datasets of real financial transactions and by exploiting the transferability property of AML, Carminati et al. achieved evasion rates ranging from 60 to 100\% with the target fraud detector considered.
Qingyu et al.~\cite{DBLP:conf/www/GuoLAHHZZ19} also analyzed the vulnerability of deep learning based fraud detectors to adversarial attacks.
They modeled the adversarial example crafting process as a search and optimization problem, presenting new algorithms inspired by the iterative FGSM. 
They conducted their experiments on Taobao's fraud detector, impairing its performance dramatically, as the precision decreased from nearly 90\% to as low as 20\%.

While interesting conclusions can be made from these studies regarding the application of AML in the fraud detection domain, none of them considered the fact that the success of the fraudulent transaction generated using an adversarial attack (i.e., the ability of the transaction generated to evade detection by the fraud detector) depends on the specific user account compromised; thus, the studies do not assess the fraud detection system's performance for different target users given each user's unique benign behavior.
In this study, we introduce and explore users' fraudability, i.e., susceptibility to fraud as reflected by the extent to which the fraud detection system can detect fraud in the user's account.
In addition, inspired by previous studies, we propose a technique for crafting adversarial (fraud) transactions.
With this technique and the approach we propose, we fill in the existing research gap in the fraud detection domain.
\section{\label{sec:approach}Proposed Approach}

\label{sec:proposed_approach}

\begin{figure*}[ht!]
\includegraphics[width=\textwidth]{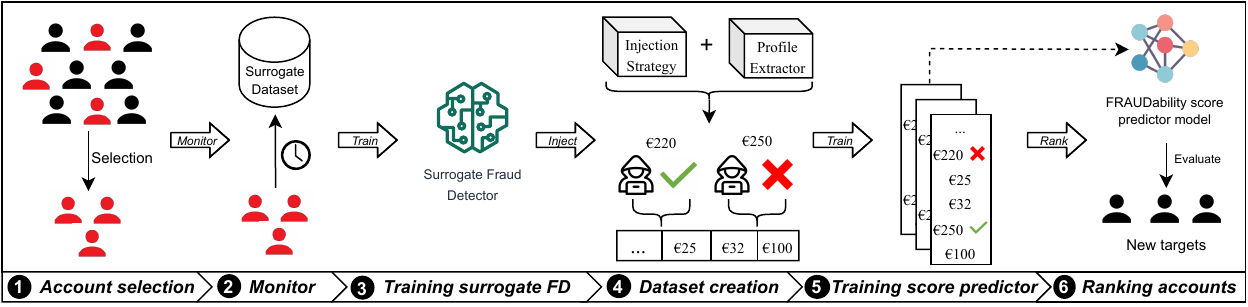}
\caption{Overview of the proposed method.}
\label{fig:method}
\end{figure*}

In this section, we present our proposed approach and describe how \method estimates fraud detection systems' performance for every user.
\method is a multi-stage approach, where each stage in the pipeline (illustrated in Figure~\ref{fig:method}) depends on the stages that precede it.
\newline

\subsection{Notations}

We define the following notations:

\begin{itemize}
    \item $u \in U$ - a user account in the financial platform ($U$ is the set of all available user accounts in the platform).
    \item $U^* \subset U$ - a subset of selected user accounts that is monitored, logged, and used to train the fraudability score prediction model.
    \item $T^u = \{t^u_1, t^u_2, ... , t^u_{|T^u|}\}$ - a set (sequence) of financial transactions performed by $u \in U$.
    \item $\mathcal{D}_{U} = \{T^u \mid u \in U \}$ - a dataset containing the transactions of each user account $u \in U$.
    \item $\mathcal{FD}$ - the fraud detector used by the financial platform to detect fraudulent transactions.
    \item $\mathcal{\hat{FD}}$ - a surrogate fraud detector used to classify sequences of transactions as fraudulent or benign.
    \item $\mathcal{P}_u$ - a set of features that represent the profile of user $u \in U$.
    \item $s_u$ - the fraudability score of user $u \in U$.
    \item $H(\mathcal{P}_u) = \hat{s_u}$ - the fraudability score predicted by the regression model $H$ for user $u \in U$.
    \item $\mathcal{S} \in \{\text{Pure Random, Predicted Amount, Late Detection},\\ \text{Max Profit}\}$ - the injection strategy used to craft and inject adversarial transactions.
    \item $n$ - the transaction sequence size based on which we calculate the features for making a prediction and performing our analysis. 
    Based on the preliminary experimental results, a transaction sequence size of 10 was shown to represent the time window most effectively. 
    A larger sequence size introduced unwanted noise, and shorter sequences did not have sufficient data. 
    Thus, in our evaluation we set $n = 10$.
\end{itemize}

\subsection{Account Selection for Monitoring}
In the first stage, the subset of user accounts $U^* \subset U$ is selected; the data of these users' accounts are used to generate the score prediction model.
The set  $U^*$ is selected arbitrary, however it should be a representative set of the users population.
Attackers usually do not have access to a large amount of financial data, unless they act maliciously by using a banking Trojan~\cite{continella2017prometheus} to steal user credentials and obtain access to many users' accounts.
Therefore, an attacker is more limited in his/her ability to select the number and type of user accounts that make up $U^*$.
From the defender (i.e., a financial e-commerce platform) standpoint, it is important to select a well-distributed, representative set of user accounts to allow \method to generalize to new users.
Although the defender has access to \textit{all} user accounts, in Appendix~\ref{appndix:running_times} we show that because financial e-commerce companies can have hundreds of millions of registered users, it is infeasible to derive a fraudability score for each individual user, and therefore, a subset of user accounts on which the fraudability score prediction model will be trained, is needed.

Next, in the second stage, the financial activity (transactions) $T^{u^*}$ of each $u^* \in U^*$ is monitored and logged.
The collected data contains various features, among which are the timestamp, transaction amount, and type of item purchased.
The longer users' activity is monitored, the richer and more valuable the data collection will be.
Transaction monitoring is a time-driven task, so in order to enrich the data, statistical features are calculated as well.
The output of this phase is the surrogate dataset $\mathcal{D}_{U^*}$, which is used to train the surrogate fraud detector $\hat{\mathcal{FD}}$, as will be explained in the next stage.

\subsection{Surrogate Fraud Detector}
Using the surrogate dataset $\mathcal{D}_{U^*}$, in stage 3 we train a surrogate fraud detector $\hat{\mathcal{FD}}$.
This surrogate detector is able to classify a given transaction as fraudulent or benign, similar to the fraud detector used by the service provider $\mathcal{FD}$.
As mentioned earlier, since a sequential representation of the data is used, sequential classifiers such as LSTM-based models are preferable.
Therefore, an LSTM undercomplete autoencoder (UAE) is used in this research. 
The objective of the UAE is to capture the most important features present in the data. 
The hidden layer of UAEs has a smaller dimension than the input layer.
To detect fraud (i.e., anomalies) the UAE receives a sequence of transactions $x$ of length $n$, each of which contains the feature set $W$.
The feature set $W$ contains both numerical data (e.g., the purchase amount) and categorical data (e.g., the type of item purchased).
We handle categorical data by using a feature embedding technique, which is an efficient technique for this type of data~\cite{DBLP:conf/aaai/HuangSYH19}.
The sequence $x$ is then fed into the encoder $\varphi$ to obtain the encoded representation $z$.
The encoded representation $z$ is later fed into the decoder $\psi$, resulting in the reconstructed sequence $\hat{x}$, from which the reconstruction error is calculated.
Based on this reconstruction error and a predefined threshold, anomalies (i.e., fraud) are determined.
Algorithm~\ref{algorithm:detecting_frauds} describes the process of determining anomalies using our fraud detector. 
The encoder and decoder can be defined as transitions $\phi$ and $\psi$ such that:
\begin{equation} \label{encoder}
  \varphi:\mathcal{X} \rightarrow \mathcal{F}
\end{equation}
\begin{equation} \label{decoder}
  \psi:\mathcal{F} \rightarrow \mathcal{X}
\end{equation}
\begin{equation}
  \varphi,\psi= \argmin_{\varphi,\psi} {\lVert X-(\mathbf{\varphi}\circ\psi)X \rVert}^2
\end{equation}
 
\begin{figure}[t]
\includegraphics[width=0.5\textwidth]{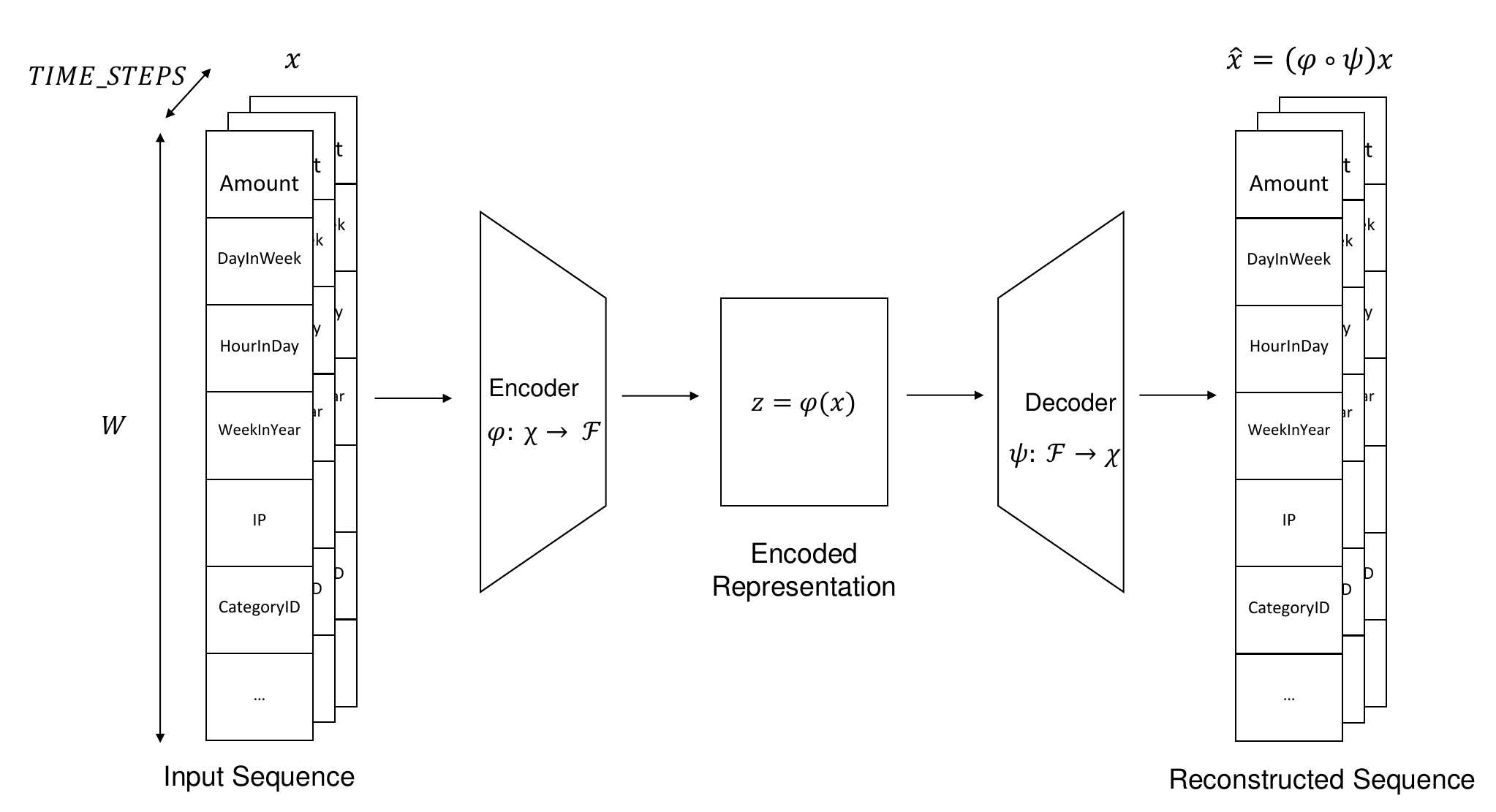}
\caption{LSTM autoencoder architecture used for fraud detection in this research. We define anomalies based on the reconstruction error. Both surrogate and target fraud detectors use this scheme.}
\label{fig:fraud-detector}
\end{figure}

Figure~\ref{fig:fraud-detector} presents the architecture of the fraud detectors used in our research.

\subsection{Dataset Creation}

In the fourth stage, we generate a training set which will be used to train the fraudability score predictor.
The training set consists of feature vectors $\mathcal{P}_u$, which are descriptive user profiles for each user $u^* \in U^*$, and a (fraudability) score (i.e., a label) $s_u$, which indicates the user's fraudabilty level.
In order to generate the training set, that is, to collect user profiles and their corresponding scores, we first inject carefully crafted fraudulent transactions into the legitimate transactions of each user $u^* \in U^*$.
We then apply $\hat{\mathcal{FD}}$ (trained in the previous stage) in order to estimate the ability of a fraud detector to detect a user's fraudulent transactions (based on the ability of $\hat{\mathcal{FD}}$ to classify the injected fraudulent transactions as anomalous).
The fraudability score is always calculated with respect to some target metric that we with to maximize. 
In other words, the definition of "fraudable" is somewhat subjective and is dependent on the goals of the platform owner; while one platform may benefit from protecting users who are prone to high number of frauds, others may appreciate protecting users who tend lose large amount of money, when get fraud. 
In section \ref{metrics} we present three such targets metrics ("Injection Rate", "Time-to-Detection" and "Money Stolen") and test our framework on one of them.
The score is then generated by applying a procedure that crafts adversarial transactions, injects them into a user's legitimate sequences, and checks whether sequences with the injected fraudulent transactions are detected by $\mathcal{\hat{FD}}$. 
The outcome of this stage is a training set consisting of training examples
\textlangle{}$\mathcal{P}_u$,$s_u$\textrangle{} used to train the fraudability score predictor.

We now describe the two important steps performed in this stage, namely \textit{adversarial transaction injection} and \textit{user profile creation}.

\textbf{Transaction injection \& detection times.} Algorithm~\ref{algorithm:inject-and-test} describes the procedure used to inject adversarial transactions.
First, the relevant transactions of each user $u^* \in U^*$ are fetched.
Then, an adversarial transaction is generated according to the predefined injection strategy $\mathcal{S}$.
This transaction is injected into the user's legitimate sequence. 
Next, all sequences that contain the injected transaction under test are extracted. 
Then, $\mathcal{\hat{FD}}$ predicts whether each individual extracted sequence is anomalous.
If all of the sequences containing the injected transaction examined are classified as benign, the fraud detector classifies the transaction as benign.
In all other cases is it considered fraudulent.

\begin{figure*}[t]
\includegraphics[width=\textwidth]{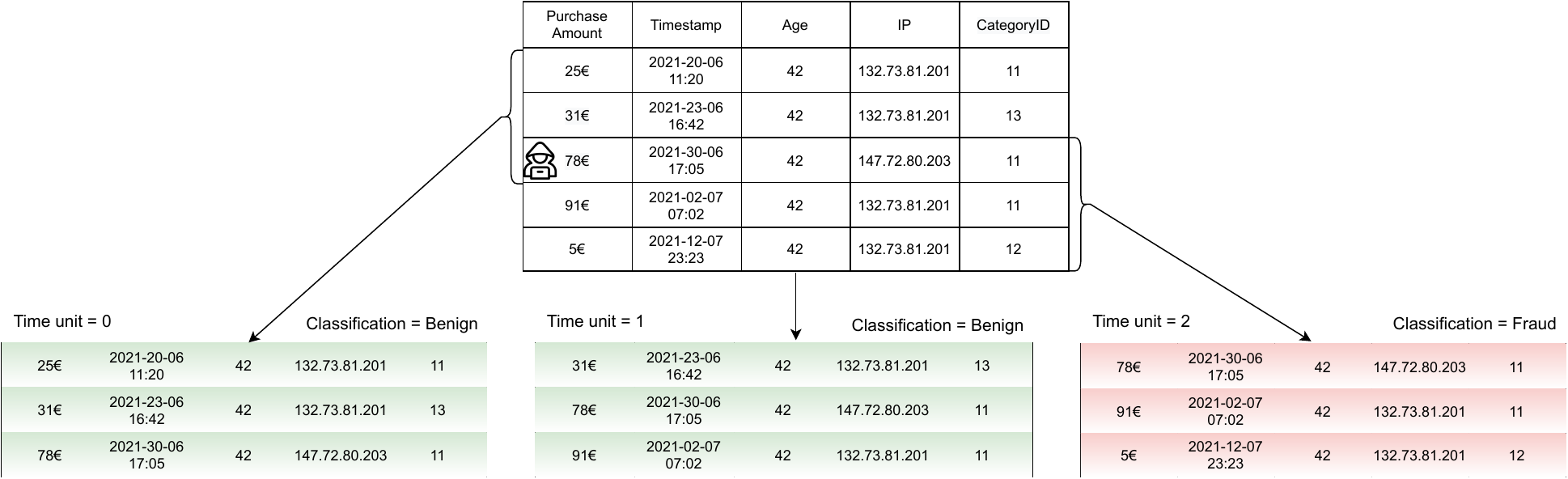}
\caption{Sequential classification approach used in this work. The middle transaction (78€) is under test. We examine each sequence that contains this transaction to determine its class - fraud or benign. The last sequence (time unit 2) is anomalous; thus the injection was detected at time unit 2 (out of two).}
\label{fig:sequential_classification}
\end{figure*}

This approach also has benefits when modeling detection times. 
The sequential approach described above allows a more in-depth discussion on the detection time of a single transaction $T$; let $n=3$ be our fraud detection time window; then each transaction inspected will appear in at most three sequences, where the first sequence (in which $T$ appears last) is considered as time unit 0, and the last sequence (in which $T$ appears first) is considered as time unit $3-1=2$. 
Figure~\ref{fig:sequential_classification} demonstrates this sequential classification approach when a transaction is detected at time unit 2 (out of two), meaning that sequences \#0 and \#1 were classified as \textit{benign}, but in the third sequence (which is sequence \#2), in the context of some future transactions, the classification will be \textit{fraud}.
We repeat this process for every possible injection location and collect the data for successful injections.
Finally, for each injection attempt, we calculate the injection metrics, namely the successful \textit{injection rate}, average \textit{time to detection}, and the total \textit{amount stolen} from the user under evaluation, and generate a fraudability score for each user $u^* \in U^*$.

Our dataset contains various features for each transaction, including features that relate to the transaction (e.g., the purchase amount, the type of item purchased) and features that relate to the user (e.g., the customer's age). 
Practically, an attacker can only directly control the following raw candidate transaction features: the purchase amount, the type of item purchased, and the temporal features (e.g., the day of the week, the hour of the day).
In order to our method to support both attackers and defenders who might implement it, we change the abovementioned features only when crafting adversarial transactions.
The process of crafting an adversarial transaction is heavily dependent on the preferred injection strategy; to generate adversarial transactions, we apply a different algorithm for each injection strategy employed (in Section~\ref{sec:adv_attack} we describe our proposed adversarial technique for crafting fraudulent transactions).

\textbf{User profile creation.} \label{customer_profile} 
In order to assign each user a valid label (i.e., a score) $s_u$, the fraudability score predictor $\mathcal{H}$ is fed a descriptive user profile $\mathcal{P}_u$ for each user $u^* \in U^*$. 
These user profiles allow us to capture the financial behavior of each individual user.
To obtain $\mathcal{P}_u$, i.e., the profile for user $u$, statistical attributes regarding the current state of the user are aggregated. 
More specifically, we calculate the mean ($\mu$), median ($\tilde{x}$), standard deviation ($\sigma$), minimum and maximum values $x_{min}$, and $x_{max}$ over the available transactions $T^{u^*}$ . 
The profile of user $u$ with respect to an input feature $j$ can thus be characterized in terms of the space $\mathcal{P}_u$, whose elements encode the abovementioned components as: $\mathcal{P}_u^j = (\mu_u^j,\tilde{x}_u^j,\sigma_u^j,x_{{min}_u}^j,x_{{max}_u}^j)$. 
These aggregations are performed for each input feature in the feature set $W=x_1,...,x_{|W|}$, yielding a profile vector of size $|W|$:
\begin{equation}
\mathcal{P}_u=
\begin{bmatrix}
\mathcal{P}_u^1 & \mathcal{P}_u^2 & \dots & \mathcal{P}_u^{|W|-1}  & \mathcal{P}_u^{|W|}
\end{bmatrix}
\end{equation}
This in turn expands to a profile matrix of size $5 \times |W|$. 
These aggregated attributes represent the behavioral patterns of the user.

\subsection{fraudability score predictor training} 
Finally, equipped with both fraudability scores and user profiles (the dataset), we are ready to train the predictor $\mathcal{H}$.
The idea is to train a regression model that will learn the mapping:
\[\mathcal{H}: \mathcal{P}_u \rightarrow s_u\]
With such a model, we will be able to produce fraudability scores for new users in the ranking accounts stage.
This score represents how susceptible the user is to fraud, as implied by his/her financial behavior.

\subsection{Ranking Accounts}
When a new user account (i.e., a new unseen user account to evaluate) is encountered, a new user profile $\mathcal{P}_j$ can be calculated.
This profile  will serve as the input to the fraudability score predictor to obtain a up-to-date fraudability score $s_j$.
This score can now be used to determine the extent to which a fraud detector will be able to detect financial fraud in this user account.

\begin{algorithm}
\caption{Inject-and-Test procedure used to injected the crafted transaction to the user's legitimate activity.
The function $injectTransaction(i,data,transaction)$ injects the \textit{transaction} into \textit{data} in the \textit{i'th} place.}
\label{algorithm:inject-and-test}
\DontPrintSemicolon
\KwIn{$\mathcal{\hat{FD}}$ is the surrogate fraud detector, $\mathrm{u}$ is the user being evaluated, $\mathit{S}$ is the injection strategy, $n$ is the time window.}
\KwOut{InjectionRate, TimeToDetection, AmountStolen}

 injections, successful, amountStolen, TimeToDetection $\gets$ 0\;
 TimeToDetection $\gets$ 0\;
 $D \gets$ fetchUserTransactions(u)\;
 \For {i $\gets$ 1 to NUM\_TRANSACTIONS-n}
 {
    transaction $\gets$ craftTransaction($\mathit{i,D,\mathcal{\hat{FD}},S})$\;
    injected $\gets$ injectTransaction(i,D,transaction)\;
    seq $\gets$ injected[i:i+n]\;
    anomaly $\gets$ $\mathcal{\hat{FD}}$(seq, n)\;
    \uIf{not anomaly}
    {
        successful += 1\;
        injections += 1\;
        AmountStolen += transaction.amount\;
        TimeToDetection += transaction.detectionTime\;
    }
    \Else
    {
    injections += 1\;
    }
}
InjectionRate $\gets \dfrac{successful}{injections}$\;
TimeToDetection $\gets \dfrac{TimeToDetection}{injections}$ \;
 \Return {(InjectionRate,TimeToDetection,AmountStolen)}

\end{algorithm}

\SetKwBlock{Repeat}{repeat}{}
\begin{algorithm}
\caption{Reconstruction error-based fraud detection algorithm}
\label{algorithm:detecting_frauds}
    \DontPrintSemicolon
    \KwIn{Sequences of transactions to classify $x_1,x_2,...,x_n$}
    \KwOut{Fraud or benign values bases on reconstruction error $L(x,\hat{x})$}
    
    $\phi, \psi \gets$ trained network parameters using Equations \ref{encoder} and \ref{decoder}\;
    $\alpha \gets$ threshold of reconstruction error\;
    \Repeat{
        \For{$i \gets 1$ to $n$}{
            Compute reconstruction error $L(x_i,\hat{x_i})$\;
            $L(\phi,\psi;t_i)={\lVert(x_i-(\mathbf{\phi}\circ\psi)x_i)\rVert}^2$\;
            \uIf{$reconstruction\:error$ $L(x_i,\hat{x_i}) > \alpha$}
            {
                $x_i$ is fraudulent\;
            }
            \Else
            {
                $x_i$ is legitimate\;
            }
            \label{endfor}
        }
    }
\end{algorithm} 
\section{Proposed AML Attack} \label{sec:adv_attack}

\begin{figure*}[ht]
\includegraphics[width=\textwidth]{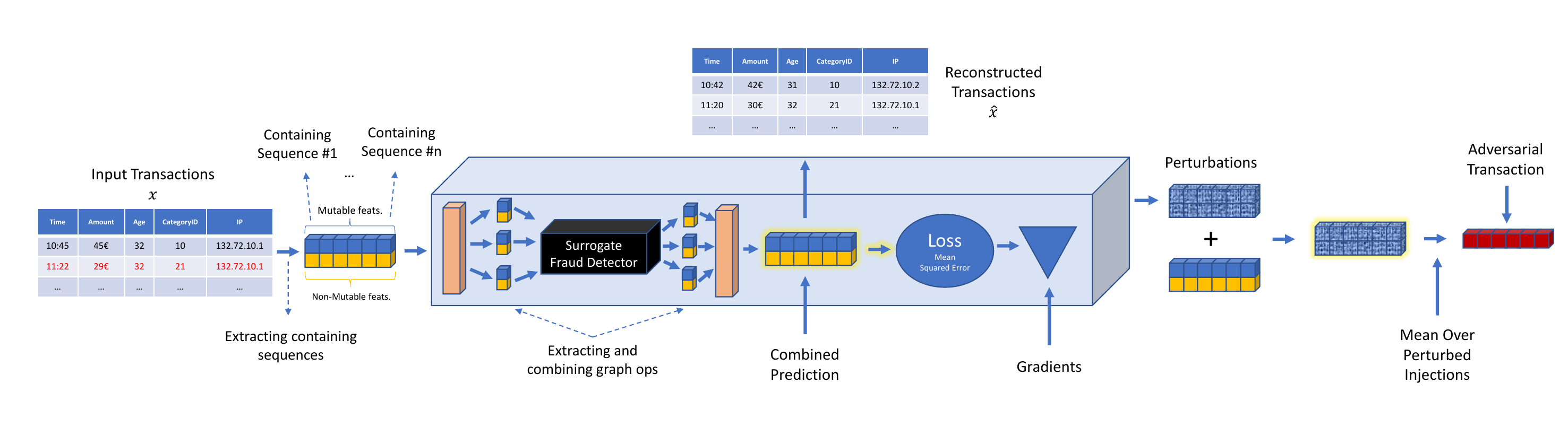}
\caption{An overview of the the adversarial transaction crafting process.
Both optimization strategies rely on this pipeline.
The colors blue and orange are respectively mutable and non-mutable features. Every cube represents a sequence.
The tall orange rectangle is a sequence of all extracted sequences that contain the injected transaction.}
\label{fig:adv-atk}
\end{figure*}

In this section, we describe our proposed technique for crafting fraudulent transactions (stage 4 in Figure~\ref{fig:method}) using an adversarial machine learning (AML) approach.
This technique is one of the contributions of our work and serves as a means of generating the fraudability scores.
We use this AML technique in stage 4 of our proposed method (see Section~\ref{sec:approach}) using two possible injection strategies, namely \textit{late detection} and \textit{profit maximization}, and compare them with two simpler injection strategies (\textit{pure random} and \textit{predicted amount}) discussed in Section~\ref{subsub:strategies}.
It should be noted that recent work on AML has focused mainly on neural networks and domains where they excel (such as computer vision), while constructing an attack on models with heterogeneous input spaces and tabular data is challenging.
We propose an attack on tabular data, which can be implemented with the FGSM attack of Goodfellow et al.~\cite{DBLP:journals/corr/GoodfellowSS14} or by its extension, the BIM attack of Kurakin et al. \cite{DBLP:conf/iclr/KurakinGB17a}.
We chose the FGSM and BIM methods as our base algorithm, because they are easy to implement (i.e., by an arbitrary attacker) and promises adequate performance.

\textbf{Attack pipeline.} As shown in Figure~\ref{fig:adv-atk}, the adversary starts from a sequence which consists of $n$ transactions.
The injected transaction is an initial candidate transaction, which is the transaction the attacker wishes to execute on behalf of the victim, without getting detected.
The attack consists of the following steps:
\begin{enumerate}
    \item Inject (i.e., insert) the transaction into a sequence of the user's legitimate transactions (shown in red).
    \item Extract all sequences containing the injected transaction.
    \item Classify the sequences using the surrogate fraud detector (a derivative-able model $\mathcal{\hat{FD}}$) to obtain the reconstructed sequences.
    \item Calculate the loss and compute the gradients of the loss function with respect to the extracted sequences. 
    \item Use the sign of the gradient (\verb|tf. sign|) to calculate the perturbations, which will be used to distort the original transaction.
    \item Update the mutable features of the transaction (in all sequences it appears in) in a direction (add or subtract) directed by the perturbation, with different values of $\epsilon$ controlling the magnitude of the distortion.
    At the end of this step, we end up with $n$ different sequences; in each sequence the injected transaction is perturbed slightly differently than the others. 
    \item Calculate a mean over those slightly-modified injected transactions to obtain the adversarial transaction.
\end{enumerate}

Algorithm \ref{algorithm:adversarial_strategies} describes the adaptation of FGSM/BIM used in this work.

\begin{algorithm}
    \caption{Adversarial transaction crafting algorithm.}
\label{algorithm:adversarial_strategies}
    \DontPrintSemicolon
    \KwIn{Attacker's transaction $t$, historical transactions $C$, the function learned by the fraud detector during training $F$, loss function used by the fraud detector $L$, $\Delta$ is either LATE-DETECT or MAX-PROFIT and $ALG$ is either FGSM or BIM, $\alpha$ is the BIM step size and was set to 1 across all experiments.}
    \KwOut{Adversarial transaction adv\_trans}
    $perturbations = []$\;
    $injected \gets injectTransaction(C, t)$\;
    $sequences \gets extractSequences(injected)$\;
    $Y^* \gets F(sequences)$\;
    $G = \nabla L(Y^*)$\;
    $S = sign(G)$\;
    \ForEach {$\epsilon \in [0.1,0.2,...,0.9]$}
    {
        \If{$ALG$ == FGSM}
            {
                $sequences^* = sequences - sequences \cdot \epsilon \cdot S$\;
            }
        \If{$ALG$ == BIM}
            {
                $sequences^*_{0} = sequences$\;
                \ForEach {$i \in [0,...,\min(\epsilon + 4, 1.25\epsilon)]$}
                    {$sequences_{i+1}^* = sequences^*_i - sequences_i \cdot \alpha \cdot S$\;}
            }
     
     candidate = mean([$sequences^*$[i,n+i,:]] for i in range(n)])\;
     candidates.append(candidate)\;
     }
    
    \uIf{$\Delta$ == MAX-PROFIT}
            {
                adv\_trans = sort(candidates, by=PurchaseAmount)\;
            }
    \Else
    {
        $detectionTimes = FD(candidates)$\;
        adv\_trans = sort(candidates, by=detectionTimes)\;
    }
    
    \Return {adv\_trans}
    
\end{algorithm}

When executing fraud transactions with the AML technique, two different optimization strategies can be utilized: \textit{profit maximization} and \textit{late detection}.

In the profit maximization strategy, the optimization process focuses on generating a large number of fraudulent transactions (while risking early detection).
This optimization strategy is implemented by selecting the perturbation (i.e., sequence of transactions) for which the largest amount of money will be stolen (obtained from undetected transactions) from all of the perturbations created with different values of $\epsilon$.
In the late detection strategy, the attacker aims to stay undetected for a long period of time (while compromising on the amount stolen).
This optimization strategy is implemented by selecting the perturbation for which the detection time is the latest, from all of the perturbations created with different values of $\epsilon$.
\section{Experimental Evaluation} \label{sec:experimental_evaluation}

In this section, we present the results of the extensive experiments conducted to evaluate the performance of our proposed method and showcase how fraudability scores helps to estimate the performance of a AI-based fraud detection system.

\subsection{Evaluation Goals}
Our experimental evaluation was conducted \textit{from an attacker’s point of view}.
The goal of the evaluation is to understand the extent to which the ability to predict users' fraudability scores can help the attacker estimate the performance of a fraud detector for a given user.  

In order to conduct the evaluation, we implemented the method (described in Section~\ref{sec:approach}) and evaluated it using a real e-commerce platform dataset.
The  evaluation aimed to answer the following research questions:
\begin{itemize}
  \item \textbf{RQ\#1} How accurate is \method's score predictor in predicting the fraudability scores of arbitrary users.
  \item \textbf{RQ\#2} Is it more beneficial to use \method's score predictor as a preliminary step of an attack (i.e., does it help the attacker determine which compromised user accounts to attack) than to perform a regular attack in which the randomly chosen compromised accounts are attacked.
  \item \textbf{RQ\#3} What is the performance of our proposed adversarial attack on sequential fraud detectors.
\end{itemize}

The first research question can be answered by using metrics like the mean square error (MSE) on an independent test set of users.
The second and third research questions can be answered through extensive experiments on a test set of users.
In these experiments, we simulate an attacker that has obtained control of a number of user accounts in a scenario in which the attacker uses \method as a preliminary step before launching the attack and compare the results of such an attack to an attack in which no such step is taken.

\subsection{Threat Model} \label{subsec:threat_model}
Defining the threat model is an essential step in designing an attack. 
Therefore, we discuss the \textit{attacker's goal, knowledge,} and \textit{capability} regarding the attack. 

In our threat model, the attacker performs an attack that violates the \textit{integrity} and \textit{confidentiality} of the fraud detector as follows.
The attacker aims to identify fraudable users and attack them; in this way the attacker exposes "weak spots" of the detector, thereby compromising the model's integrity.
In addition, latent information about the model is exposed (i.e., information regarding who is more fraudable with respect to the detector), and thus the \textit{confidentiality} is also considered compromised.

We define our attacker's knowledge based on the characterization of Biggio et al.~\cite{biggio2018wild}. 
According to the authors, an attacker can have different levels of knowledge of the targeted system, including: knowledge of the training data $\mathcal{D}$, the feature set $\mathcal{X}$, the learning algorithm $f$, along with the objective function $\mathcal{L}$ minimized during training, and possibly its (trained) parameters/hyperparameters $w$. 
The attacker's knowledge can thus be characterized in terms of the space $\Theta$ whose elements encode the abovementioned components as $\theta=(\mathcal{D},\mathcal{X},f,w)$. 
Our attacker is assumed to know the feature representation $\mathcal{X}$ in order to be able to train the fraudability score predictor, as explained in Section~\ref{sec:approach}. 
Our attacker also knows the kind of learning algorithm $f$ used (i.e., whether the classifier is linear or an unsupervised learning autoencoder model). 
Our attacker \textbf{is not} assumed to know the target model's training data $\mathcal{D}$ or the target model's (trained) parameters $w$.
By adapting the terminology from Biggio et al., our attack is thus considered a \textbf{limited knowledge (LK), gray-box attack}. 
In our attack methodology, the attacker is, however, assumed to be able to collect a surrogate dataset $\hat{\mathcal{D}}$ \footnote{We use the \textit{hat} notation to denote partial knowledge of a given component.} from a similar source (i.e., store or platform), ideally with a similar underlying data distribution.
This enables our attacker to estimate the model's parameters $\hat{w}$ from $\hat{\mathcal{D}}$, by training the \textit{surrogate fraud detector}. 
After further adaptation of the authors' characterization of the attacker's knowledge, our attack is considered a \textbf{limited knowledge attack with surrogate data (LK-SD)} and is denoted as $\theta_{LK-SD}=(\hat{\mathcal{D}},\mathcal{X},f,\hat{w})$.

With respect to the attacker's capabilities, we assume that the attacker is able to: (1) compromise users' accounts (e.g., with malware), (2) execute transactions on behalf of the users, and (3) retrieve/observe the transactions performed by the users.

\subsection{Dataset and Privacy}

Our dataset comes from a well-known e-commerce platform.\footnote{Our dataset is proprietary and cannot be disclosed due to privacy issues.} 
The dataset was collected between the years 2010 and 2012 and contains 3,219,985 real transactions performed by 556,256 customers.
In this work and due to privacy concerns, we only used a limited set of features: the purchase amount, the unique ID associated with the user; age; payment type; the date and time the purchase was executed; the hashed IP address of the connection from which the transaction was performed; and the category ID of the product purchased.
These features, to our humble opinion, are sufficient to describe customer's spending habits.

Fully aware of the ethical issues of using real-world financial data, we have taken serious measures steps to ensure privacy.
The dataset is fully anonymized and does not include any PII; 
\begin{enumerate}
  \item Sensitive features like user names, addresses and geographic location were removed by a third party before the authors received the data.
  \item The only purchase related feature are amount, category ID and the timestamp.
  \item IP addresses were hashed using SHA-3 algorithm.
\end{enumerate}

One important aspect to address is the split of the data to train and test sets.
In this work, we used used the \textit{user} as the basic data point to split on -- we used train customers and test customers whose transactions make the train set and test set, respectively.
We also validated that the train and test sets have similar distributions of customers in terms of spending habits.

\subsection{Evaluation Setup} \label{sub:evaluation_setup}

In this section, we describe the evaluation setup used to test the validity of our method. More specifically, we describe the models used and the techniques to inject synthetic fraudulent transaction.

\subsection{Models}

\textbf{Target (attacked) fraud detection model.} The target fraud detector $\mathcal{FD}$ is a One-class SVM (OC-SVM) model or a Local outlier factor (LOF) models.
The OC-SVM model was constructed with a RBF kernel and a degree of the polynomial kernel function of 3.
The LOF model with a number of neighbors of 20, leaf size of 30 and a minkowski distance computation function. 
The choice of those models was due to the fact that they were proven to be reliable and robust unsupervised anomaly detection algorithms \cite{DBLP:conf/ant/XuYLL13}.
The target fraud detector was trained on different data than the data used to construct the adversarial transactions.

\noindent \textbf{Surrogate fraud detection model.} Our surrogate fraud detector $\mathcal{\hat{FD}}$ is an LSTM undercomplete autoencoder (UAE) with three hidden layers with 256, 64, and 256 units and a tanh (hyperbolic tangent) function as the activation function. 
We used the Adam optimizer as our optimizer and the mean square error (MSE) as our loss function.

\noindent \textbf{Fraudability score model.} The fraudability score predictor can be implemented using various ML regression models, which are the algorithms most used in the literature~\cite{mercer1990fraud, wilson2009analytical, sahin2011detecting}.
These models were used to map a user profile $\mathcal{P}_u$ to its corresponding fraudability score $s_u$.
In our work, we used the following models for this task: artificial neural network (with three hidden layers with 64, 32, and 4 units and a ReLU activation function, and an output layer with a linear activation function; the loss function used was the mean absolute error (MAE) with the Adam optimizer)~\cite{DBLP:conf/ann/1995}, random forest regression (with 100 estimators and a maximum depth of 14)~\cite{liaw2002classification}, XGBoost regression (with 200 estimators and a maximum depth of 20, with a learning rate of 0.1)~\cite{DBLP:conf/kdd/ChenG16}, and linear regression (with an L2 regularization with parameter $C=\frac{1}{\lambda}=10$).

\subsection{Injection Strategies} \label{subsub:strategies}
Attackers can have different goals and wishes regarding their compromised user account: while one attacker might want to remain undetected as long as possible, potentially limiting the amount of money stolen, another might want to steal as much money as possible, as quickly as possible. 
In order to support this idea, we consider four injection strategies by which an arbitrary attacker can inject the adversarial transaction(s). 
Evaluating different injection strategies is not only useful in the context of different types of attackers, it is also useful as a precaution for the attacker himself; by examining various types of strategies,  an attacker can choose the strategy that best meets his/her objectives and priorities.

\noindent \textbf{Pure random.} This strategy represents the attacker with the least amount of knowledge and serves as a baseline for other, more sophisticated injection strategies. 
In this strategy, an attacker that has gained control of a user account will randomize the transaction's mutable features, hoping to remain undetected by the fraud detector.\footnote{We set a random seed in our implementation for reproducibility.}

\noindent \textbf{Predicted amount.} In this strategy, the attacker trains a deep LSTM network, a Forecaster.
The Forecaster is a long short-term memory artificial neural network (LSTM ANN) that aggregates historical user transactions and models the user's spending patterns. 
Given some past user transactions, it predicts the amount of the user's next transaction.
Having predicted that, the attacker then crafts an adversarial transaction of that amount. 
This strategy represents a smarter attacker that tries to "guess" the next transaction of the user based on the previous ones.
We implemented the Forecaster as a sequence-to-value LSTM network which consists of two hidden layers with 128 LSTM blocks and the tanh activation function, each. 
The output's layer activation function is linear.
We used the mean absolute error (MAE) as our loss function and the Adam optimizer.

\noindent \textbf{Adversarial attack strategies.} In addition to the abovementioned injection strategies, we developed two additional injection strategie: \textit{profit maximization} and \textit{late detection}.
These strategies exploit adversarial machine learning techniques to craft fraudulent transactions.
These strategies are discussed in detail in Section~\ref{sec:adv_attack}.

\subsection{Evaluation Process}

FRAUDability's performance might vary depending on the knowledge of the attacker/defender, and the knowledge of the former is tightly coupled with the amount of data he/she obtains.
Also, in order to achieve statistical significance to our tests, we filtered users with less than 30 transactions, ending up with a total of 6,405 users with an average of 32.45 transaction per user.
Therefore, we tested our framework on different settings based on the number of users whose data is obtained: 1,921, 1,281, and 320 users, which respectively represent 30\%, 20\%, and 5\% of the users available in our dataset (a total of 6,405 users).
In addition, we also tested the framework from the defender's perspective, which holds the entire data in his organization (6,405 users).
It is important to note that the attacker and defender are verified to have disjoint sets of users' data.
In each experiment, we calculated fraudability scores for all users, and then sorted the scores from highest to lowest.
\textbf{We then injected transactions into the top 100 most fraudable accounts, i.e., the 100 accounts with the highest fraudability scores}.
Then, the transactions were classified by the defender's fraud detector $\mathcal{FD}$.
In each experiment, we present the performance of the various type of injections on users with high fraudability scores, compared to a random selection of customers.\footnote{The results of the injection using pure random were calculated after averaging the performance obtained in 10 experiments performed with injection based on random selection.} 

\subsection{Evaluation Metrics} \label{metrics}

We evaluate the performance of our proposed method using three metrics: \textit{injection rate}, \textit{money stolen}, and \textit{time-to-detection}. 

\textbf{Injection rate.} This metric indicates the percentage of undetected injected fraudulent transactions (i.e., false negatives) in the user's account in relation to the number of injected fraudulent transactions overall.
The surrogate fraud detector $\mathcal{\hat{FD}}$ decides whether or not a fraudulent transaction is likely to be undetected.
This metric depends on the threshold of the surrogate fraud detector -- the lower the anomaly threshold we set, the lower the injection rate.

\textbf{Time-to-detection.} This metric represents the time that elapses before detection, in terms of the number of transactions (we measure time in units the amount of transactions). 
Because our fraud detector inspects sequences of transactions, a more in-depth inspection of the fraud detector's classification can be performed; since each transaction is present in at most $n$ sequences, when $n$ is our time window, we can now relate to when the fraud was detected by looking at the first sequence (containing the injected transaction) that was classified as anomaly. 
The best-case scenario for an attacker is when all of the sequences are classified as benign, while the worst-case scenario is when the first sequence (time unit 0) is classified as fraud.

\textbf{Money stolen.} This metric represents the amount of money, in euros (€), that the user loses.
This metric is heavily dependent on the two previous metrics and the injection strategy chosen. 
In addition, it provides an idea of the real-world financial impact that such attacks can have on real e-commerce platforms.

Following are the mathematical formulations of the abovementioned metrics:
Let $N$ be the number of targeted users, $F_i$ be the number of fraudulent transactions the attacker wants to perform against the i'th user, $K_{i,j}$ be the amount of money for the i'th fraudulent transaction injected into the account of the i'th user, $X_{i,j}$ be the binary variable indicating whether the j'th transaction of the i'th user was detected by the surrogate fraud detector or not, and $T_{i,j}$ be the time the j'th transaction of the i'th user was detected (or $n$ if undetected), we define the abovementioned metrics as follows:

\begin{align*}
& Injection\; rate = 1/N\cdot\sum_{i=1}^{N} \sum_{j=1}^{F_i} \neg X_{i,j}/F_i \\
& Time\text -to\text -detection = 1/N\cdot\sum_{i=1}^{N}\sum_{j=1}^{F_i} T_{i,j}/F_i \\
& Money\; stolen = 1/N\cdot\sum_{i=1}^{N} \sum_{j=1}^{F_i}\neg X_{i,j} \cdot K_{i,j}
\end{align*}

\subsection{Evaluation Results}

In this section we present the evaluation results with regard to the research questions presented earlier.

\subsection{Fraudability score predictor accuracy}

Table~\ref{table:fraudability_predicting} presents the results of the experiments conducted to evaluate the accuracy of the fraudability score predictor $\mathcal{H}$.
In each experiment, we trained a different regression model according to the setup discussed in Section~\ref{sub:evaluation_setup}.
Each evaluated model $\mathcal{H}$ was trained on the same training set of selected users $U^*$ (obtained in stage 1 of the method).
In our method, any of the three metrics provided can serve as the target variable (i.e., the fraudability score).
In our experiments, we chose the amount of money stolen from the user (scaled to [0,1]) as the target variable, although other approaches can be taken.
Our decision stems from the need to reflect real-life attacker that wishes to maximize his financial gain.
The experiments were performed on a independent test set of users $U \setminus U^*$ that were not selected in stage 1 of the pipeline, the model was not trained on and they were validated to come from similar distribution to that of the selected users $U^*$.
This experiment is performed for each of the injection strategies discussed in Section~\ref{subsub:strategies}.
From the results presented in the table, we can conclude that the fraudability score predictor is able to accurately predict the fraudability of new, unknown users.
Another conclusion that can be drawn from the results is that the random forest regression model outperforms the other regression models on this task with all injection strategies examined, with an average error of 7.25.
With regard to research question \#1, we can conclude that the fraudability score predictor is accurate to a decent rate.

\begin{table}[ht]
\centering
\caption{\method's performance (XGBoost) in the task of predicting fraudability scores. The actual values were collected using Algorithm ~\ref{algorithm:inject-and-test}. The fraudability score considered here is the amount stolen from the user's account.\\
}
\begin{tabular}{ccc}
\hline
\multicolumn{3}{c}{Injection Strategy: Pure Random}       \\ \hline
Model                   & Actual & FRAUDability \\
Neural Network          & 226.33 & 250.43       \\
Random Forest Regressor & 224.32 & 227.54       \\
XGBoost Regressor       & 241.45 & 230.21       \\
Linear Regression       & 240.97 & 201.55       \\ \hline
\multicolumn{3}{c}{Injection Strategy: Predicted Amount}  \\ \hline
Model                   & Actual & FRAUDability \\
Neural Network          & 134.45 & 179.97       \\
Random Forest Regressor & 141.24 & 130.24       \\
XGBoost Regressor       & 149.11 & 110.14       \\
Linear Regression       & 240.97 & 201.55       \\ \hline
\multicolumn{3}{c}{Injection Strategy: Adversarial Late Detection}  \\ \hline
Model                   & Actual & FRAUDability \\
Neural Network          & 274.90 & 220.24       \\
Random Forest Regressor & 220.04 & 222.87       \\
XGBoost Regressor       & 270.89 & 241.66       \\
Linear Regression       & 240.97 & 201.55       \\ \hline
\multicolumn{3}{c}{Injection Strategy: Adversarial Max. Profit}  \\ \hline
Model                   & Actual & FRAUDability \\
Neural Network          & 397.45 & 360.45       \\
Random Forest Regressor & 345.66 & 340.74       \\
XGBoost Regressor       & 374.85 & 322.54       \\
Linear Regression       & 340.97 & 301.55      
\end{tabular}
\label{table:fraudability_predicting}
\end{table}

\begin{figure}[t]
\includegraphics[width=0.5\textwidth]{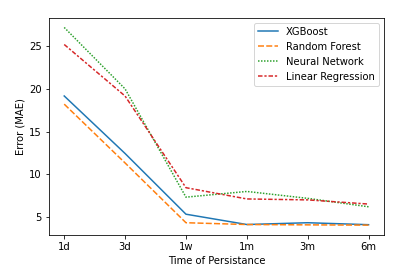}
\caption{The error of the fraudability score predictor as a function of the amount of time the attacker spends in the monitoring stage. The x-axis represents different time periods: 1 day worth of transactions, three days, etc.}
\label{fig:persistence}
\end{figure}

Another interesting issue that was investigated is the accuracy (i.e., error) of the fraudability score predictor itself, as a function of the time the attacker spends in the monitoring stage (i.e., logging user's transactions).
As shown in Figure~\ref{fig:persistence}, it can be seen that using only one week of monitoring is enough to stabilize the error of the fraudability score predictor at a reasonable rate.
After that, additional monitoring does not improve the model's performance significantly.

\subsection{Effectiveness of the fraudability score predictor}

\begin{table*}[ht]
\footnotesize
\caption{Experimental results for all injection strategies, attacker knowledge levels, and FRAUDability implementations, on a OC-SVM target (attacked) fraud detector. The best results from the attacker's points of view appear in {\color[HTML]{009901} green}, and the worst results from the attacker's point of view appear in \textcolor{red}{red}. The values are averaged over the number of users evaluated. Adversarial Max Profit and Adversarial Late Detection results are formatted as \textless FGSM\textgreater |\textless BIM\textgreater.
}
\resizebox{\textwidth}{!}
{
\begin{tabular}{ccccccc}
\multicolumn{1}{l}{}                                                                                                 &                              & \multicolumn{4}{c}{FRAUDability}                                                                                                                                              & Random Selection     \\ \hline
\multicolumn{1}{c|}{}                                                                                                & Metric/Model                 & Neural Network                            & RandomForest Regressor                    & XGBoost Regressor                         & Linear Regression                         & \multicolumn{1}{l}{} \\ \hline
\multicolumn{7}{c}{Strategy: Pure Random}                                                                                                                                                                                                                                                                                                                  \\ \hline
\multicolumn{1}{c|}{}                                                                                                & Injection Rate               & 40.14\%                                   & {\color[HTML]{009901} 42.84\%}            & 36.73\%                                   & {\color[HTML]{FE0000} 33.62\%}            & 24.33\%              \\
\multicolumn{1}{c|}{}                                                                                                & Time-to-Detection {[}0-10{]} & 6.46                                      & {\color[HTML]{009901} 5.74}               & 5.32                                      & {\color[HTML]{FE0000} 4.47}               & 2.40                 \\
\multicolumn{1}{c|}{\multirow{-3}{*}{\begin{tabular}[c]{@{}c@{}}Defender's \\ Knowledge:\\ 6400 users\end{tabular}}} & Money Stolen                 & € 780                                     & {\color[HTML]{009901} € 789}              & € 637                                     & {\color[HTML]{FE0000} € 634}              & € 215.37             \\ \hline
\multicolumn{1}{c|}{}                                                                                                & Injection Rate               & 36.82\%                                   & {\color[HTML]{009901} 38.44\%}            & {\color[HTML]{000000} 34.37\%}            & {\color[HTML]{FE0000} 31.48\%}            & 24.33\%              \\
\multicolumn{1}{c|}{}                                                                                                & Time-to-Detection {[}0-10{]} & 4.46                                      & {\color[HTML]{009901} 4.92}               & 4.21                                      & {\color[HTML]{FE0000} 3.74}               & 2.40                 \\
\multicolumn{1}{c|}{\multirow{-3}{*}{\begin{tabular}[c]{@{}c@{}}Attacker's \\ Knowledge:\\ 1921 users\end{tabular}}} & Money Stolen                 & € 566                                     & {\color[HTML]{009901} € 574}              & {\color[HTML]{000000} € 551}              & {\color[HTML]{FE0000} € 542}              & € 215.37             \\ \hline
\multicolumn{1}{c|}{}                                                                                                & Injection Rate               & {\color[HTML]{009901} 34.21\%}            & {\color[HTML]{000000} 32.41\%}            & {\color[HTML]{000000} 33.33 \%}           & {\color[HTML]{FE0000} 30.59\%}            & 24.33\%              \\
\multicolumn{1}{c|}{}                                                                                                & Time-to-Detection {[}0-10{]} & 4.02                                      & {\color[HTML]{009901} 4.14}               & 3.22                                      & {\color[HTML]{FE0000} 3.12}               & 2.40                 \\
\multicolumn{1}{c|}{\multirow{-3}{*}{\begin{tabular}[c]{@{}c@{}}Attacker's\\ Knowledge: \\ 1281 users\end{tabular}}} & Money Stolen                 & {\color[HTML]{009901} € 543}              & {\color[HTML]{FE0000} € 517}              & € 534                                     & € 541                                     & € 215.37             \\ \hline
\multicolumn{1}{c|}{}                                                                                                & Injection Rate               & {\color[HTML]{009901} 32.47\%}            & 29.49\%                                   & 30.32\%                                   & {\color[HTML]{FE0000} 28.95\%}            & 24.33\%              \\
\multicolumn{1}{c|}{}                                                                                                & Time-to-Detection {[}0-10{]} & {\color[HTML]{009901} 3.72}               & 3.64                                      & 2.44                                      & {\color[HTML]{FE0000} 2.37}               & 2.40                 \\
\multicolumn{1}{c|}{\multirow{-3}{*}{\begin{tabular}[c]{@{}c@{}}Attacker's \\ Knowledge: \\ 320 users\end{tabular}}} & Money Stolen                 & {\color[HTML]{009901} € 541}              & € 454                                     & {\color[HTML]{FE0000} € 428}              & {\color[HTML]{333333} € 467}              & € 215.37             \\ \hline
\multicolumn{7}{c}{Strategy: Predicted Amount}                                                                                                                                                                                                                                                                                                             \\ \hline
\multicolumn{1}{c|}{}                                                                                                & Injection Rate               & 62.91\%                                   & {\color[HTML]{009901} 68.11\%}            & {\color[HTML]{FE0000} 54.62\%}            & 65.54\%                                   & 36.00\%              \\
\multicolumn{1}{c|}{}                                                                                                & Time-to-Detection {[}0-10{]} & 7.01                                      & {\color[HTML]{009901} 7.47}               & 6.45                                      & {\color[HTML]{FE0000} 5.98}               & 3.55                 \\
\multicolumn{1}{c|}{\multirow{-3}{*}{\begin{tabular}[c]{@{}c@{}}Defender's \\ Knowledge:\\ 6400 users\end{tabular}}} & Money Stolen                 & € 645                                     & {\color[HTML]{009901} € 792}              & {\color[HTML]{333333} € 560}              & {\color[HTML]{FE0000} € 541}              & € 244.67             \\ \hline
\multicolumn{1}{c|}{}                                                                                                & Injection Rate               & 56.74\%                                   & {\color[HTML]{009901} 58.21\%}            & {\color[HTML]{FE0000} 49.22\%}            & 55.54\%                                   & 36.00\%              \\
\multicolumn{1}{c|}{}                                                                                                & Time-to-Detection {[}0-10{]} & 5.66                                      & {\color[HTML]{009901} 5.81}               & 5.04                                      & {\color[HTML]{FE0000} 4.77}               & 3.55                 \\
\multicolumn{1}{c|}{\multirow{-3}{*}{\begin{tabular}[c]{@{}c@{}}Attacker's \\ Knowledge:\\ 1921 users\end{tabular}}} & Money Stolen                 & € 445                                     & € 540                                     & {\color[HTML]{009901} € 542}              & {\color[HTML]{FE0000} € 449}              & € 244.67             \\ \hline
\multicolumn{1}{c|}{}                                                                                                & Injection Rate               & 54.12\%                                   & {\color[HTML]{009901} 58.05\%}            & 52.22 \%                                  & {\color[HTML]{FE0000} 49.22\%}            & 36.00\%              \\
\multicolumn{1}{c|}{}                                                                                                & Time-to-Detection {[}0-10{]} & 5.42                                      & {\color[HTML]{009901} 5.44}               & 4.84                                      & {\color[HTML]{FE0000} 4.47}               & 3.55                 \\
\multicolumn{1}{c|}{\multirow{-3}{*}{\begin{tabular}[c]{@{}c@{}}Attacker's\\ Knowledge: \\ 1281 users\end{tabular}}} & Money Stolen                 & € 522                                     & {\color[HTML]{009901} € 546}              & € 532                                     & {\color[HTML]{FE0000} € 410}              & € 244.67             \\ \hline
\multicolumn{1}{c|}{}                                                                                                & Injection Rate               & 50.20\%                                   & {\color[HTML]{009901} 55.51\%}            & 50.41 \%                                  & {\color[HTML]{FE0000} 40.24\%}            & 36.00\%              \\
\multicolumn{1}{c|}{}                                                                                                & Time-to-Detection {[}0-10{]} & 5.21                                      & {\color[HTML]{009901} 5.24}               & 4.21                                      & {\color[HTML]{FE0000} 3.22}               & 3.55                 \\
\multicolumn{1}{c|}{\multirow{-3}{*}{\begin{tabular}[c]{@{}c@{}}Attacker's \\ Knowledge: \\ 320 users\end{tabular}}} & Money Stolen                 & € 499                                     & {\color[HTML]{009901} € 510}              & € 467                                     & {\color[HTML]{FE0000} € 404}              & € 244.67             \\ \hline
\multicolumn{7}{c}{Strategy: Adversarial Late Detection}                                                                                                                                                                                                                                                                                                   \\ \hline
\multicolumn{1}{c|}{}                                                                                                & Injection Rate               & 97.77\% | 97.72\%                         & {\color[HTML]{009901} 98.71\% | 98.23\%}  & 97.04\% | 97.61 \%                        & {\color[HTML]{FE0000} 94.27\% | 94.31 \%} & 92.84\%              \\
\multicolumn{1}{c|}{}                                                                                                & Time-to-Detection {[}0-10{]} & 9.88 | 9.89                               & {\color[HTML]{009901} 9.89 | 9.88}        & {\color[HTML]{330001} 9.61 | 9.68}        & {\color[HTML]{FE0000} 9.21 | 9.30}        & 9.21                 \\
\multicolumn{1}{c|}{\multirow{-3}{*}{\begin{tabular}[c]{@{}c@{}}Defender's \\ Knowledge:\\ 6400 users\end{tabular}}} & Money Stolen                 & € 3304 | € 3309                           & {\color[HTML]{009901} € 3434 | € 3430}    & € 3231 | € 3233                           & {\color[HTML]{FE0000} € 3011 | € 3016}    & € 2532.46            \\ \hline
\multicolumn{1}{c|}{}                                                                                                & Injection Rate               & 96.67\% | 96.99 \%                        & {\color[HTML]{009901} 98.24\% | 98.91 \%} & 96.12\% | 96.63 \%                        & {\color[HTML]{FE0000} 91.14\% | 91.74 \%} & 92.84\%              \\
\multicolumn{1}{c|}{}                                                                                                & Time-to-Detection {[}0-10{]} & 9.74 | 9.74                               & {\color[HTML]{009901} 9.77 | 9.78}        & {\color[HTML]{330001} 9.54 | 9.69}        & {\color[HTML]{FE0000} 9.26 | 9.32}        & 9.21                 \\
\multicolumn{1}{c|}{\multirow{-3}{*}{\begin{tabular}[c]{@{}c@{}}Attacker's \\ Knowledge:\\ 1921 users\end{tabular}}} & Money Stolen                 & € 3087 | € 3095                           & {\color[HTML]{009901} € 3102 | € 3196}    & € 3033 | € 3047                           & {\color[HTML]{FE0000} € 2932 | € 3023}    & € 2532.46            \\ \hline
\multicolumn{1}{c|}{}                                                                                                & Injection Rate               & 95.56\% | 96.11 \%                        & {\color[HTML]{009901} 97.12\% | 98.02 \%} & 95.02\% | 96.01 \%                        & {\color[HTML]{FE0000} 90.01\% | 90.11 \%} & 92.84\%              \\
\multicolumn{1}{c|}{}                                                                                                & Time-to-Detection {[}0-10{]} & {\color[HTML]{009901} 9.74 | 9.94}        & {\color[HTML]{000000} 9.55 | 9.62}        & 9.45 | 9.51                               & {\color[HTML]{FE0000} 9.40 | 9.45}        & 9.21                 \\
\multicolumn{1}{c|}{\multirow{-3}{*}{\begin{tabular}[c]{@{}c@{}}Attacker's\\ Knowledge: \\ 1281 users\end{tabular}}} & Money Stolen                 & € 2879 | € 2911                           & {\color[HTML]{009901} € 2970 | € 3010}    & € 2899 | € 2904                           & {\color[HTML]{FE0000} € 2823 | € 2843}    & € 2532.46            \\ \hline
\multicolumn{1}{c|}{}                                                                                                & Injection Rate               & 94.24\% | 95.04 \%                        & {\color[HTML]{009901} 95.74\% | 96.11 \%} & 93.2\% | 94.17 \%                         & {\color[HTML]{FE0000} 89.78\% | 90.11 \%} & 92.84\%              \\
\multicolumn{1}{c|}{}                                                                                                & Time-to-Detection {[}0-10{]} & 9.30 | 9.45                               & {\color[HTML]{FE0000} 9.29 | 9.31}        & 9.41 | 9.50                               & {\color[HTML]{009901} 9.44 | 9.51}        & 9.21                 \\
\multicolumn{1}{c|}{\multirow{-3}{*}{\begin{tabular}[c]{@{}c@{}}Attacker's \\ Knowledge: \\ 320 users\end{tabular}}} & Money Stolen                 & € 2810 | € 2885                           & {\color[HTML]{009901} € 2875 | € 2889}    & {\color[HTML]{FE0000} € 2543 | € 2642}    & € 2677 | € 2683                           & € 2532.46            \\ \hline
\multicolumn{7}{c}{Strategy: Adversarial Max Profit}            \\ \hline                  & Injection Rate               & 94.27\% | 94.92 \%                        & 95.27\% | 96.03 \%                        & {\color[HTML]{009901} 96.82\% | 97.11 \%} & {\color[HTML]{FE0000} 92.39\% | 93.09 \%} & 87.44\%     \\
  & Time-to-Detection {[}0-10{]} & 8.37 | 8.46                               & 8.41 | 8.67                               & {\color[HTML]{009901} 9.03 | 9.13}        & {\color[HTML]{FE0000} 8.28 | 8.41}        & 7.18                 \\
\multirow{-3}{*}{\begin{tabular}[c]{@{}c@{}}Defender's \\ Knowledge:\\ 6400 users\end{tabular}}                      & Money Stolen                 & € 4004 | € 4076                           & {\color[HTML]{FE0000} € 3674 | € 3697}    & {\color[HTML]{009901} € 4212 | € 4255}    & € 3971 | € 4001                           & € 2747.48            \\ \hline
\multicolumn{1}{c|}{}                                                                                                & Injection Rate               & 91.27\% | 91.33 \%                        & 90.47\% | 91.54 \%                        & {\color[HTML]{009901} 91.48\% | 91.66 \%} & {\color[HTML]{FE0000} 85.67\% | 85.90 \%} & 87.44\%              \\
\multicolumn{1}{c|}{}                                                                                                & Time-to-Detection {[}0-10{]} & 7.15 | 7.24                               & 7.30 | 7.29                               & {\color[HTML]{009901} 7.31 | 7.40}        & {\color[HTML]{FE0000} 7.20 | 7.21}        & 7.18                 \\
\multicolumn{1}{c|}{\multirow{-3}{*}{\begin{tabular}[c]{@{}c@{}}Attacker's \\ Knowledge:\\ 1921 users\end{tabular}}} & Money Stolen                 & € 3502 | € 3562                           & {\color[HTML]{FE0000} € 3457 | € 3511}    & {\color[HTML]{009901} € 3581 | € 3602}    & € 3504 | € 3565                           & € 2747.48            \\ \hline
\multicolumn{1}{c|}{}                                                                                                & Injection Rate               & {\color[HTML]{009901} 90.72\% | 92.55 \%} & 89.51\% | 90.14 \%                        & 89.14\% | 90.11 \%                        & {\color[HTML]{FE0000} 83.75\% | 84.56 \%} & 87.44\%              \\
\multicolumn{1}{c|}{}                                                                                                & Time-to-Detection {[}0-10{]} & {\color[HTML]{FE0000} 7.01 | 7.15}        & {\color[HTML]{000000} 7.10 | 7.25}        & {\color[HTML]{FE0000} 7.99 | 8.16}        & 7.02 | 7.03                               & 7.18                 \\
\multicolumn{1}{c|}{\multirow{-3}{*}{\begin{tabular}[c]{@{}c@{}}Attacker's\\ Knowledge: \\ 1281 users\end{tabular}}} & Money Stolen                 & {\color[HTML]{000000} € 3444 | € 3563}    & € 3399 | € 3453                           & {\color[HTML]{009901} € 3482 | € 3511}    & {\color[HTML]{FE0000} € 2997 | € 3018}    & € 2747.48            \\ \hline
\multicolumn{1}{c|}{}                                                                                                & Injection Rate               & 88.72\% | 89.33 \%                        & {\color[HTML]{009901} 89.41\% | 90.31 \%} & 89.14\% | 90.42 \%                        & {\color[HTML]{FE0000} 82.99\% | 84.22 \%} & 87.44\%              \\
\multicolumn{1}{c|}{}                                                                                                & Time-to-Detection {[}0-10{]} & {\color[HTML]{009901} 7.99 | 8.32}        & {\color[HTML]{000000} 7.05 | 7.12}        & 7.87 | 8.01                               & {\color[HTML]{FE0000} 7.03 | 7.43}        & 7.18                 \\
\multicolumn{1}{c|}{\multirow{-3}{*}{\begin{tabular}[c]{@{}c@{}}Attacker's \\ Knowledge: \\ 320 users\end{tabular}}} & Money Stolen                 & € 3299 | € 3314                           & {\color[HTML]{009901} € 3348 | € 3466}    & € 3335 | € 3417                           & {\color[HTML]{FE0000} € 3240 | € 3333}    & € 2747.48            \\ \hline
\end{tabular}
}
\label{table:FRAUDability_results_1}
\end{table*}

Table ~\ref{table:FRAUDability_results_1} presents the results of the experimental evaluation of the attacks for each injection strategy (i.e., pure random, predicted amount, late detection, and profit maximization) on the OC-SVM target model.
In appendix~\ref{appndix:target_models} we present additional evaluation on Local Outlier Factor, as well as on a modified version of the Variational Autoencoder, as the target fraud detectors.

Based on these results, we draw the following conclusions: with regard to the injection rate, an improvement of 58\% (24.33\%→38.44\%) is obtained for the pure random injection strategy and 5.5\% (92.84\%→98.91\%) for the late detection strategy, compared to random selection of users (baseline).
From the money stolen perspective, the improvement is even better, with a 166\% (€215→€574) increase in the amount of money stolen with the random injection strategy and a 31\% (€2,747→€3,602) increase in the amount of money stolen with the profit maximization injection strategy.
From the time-to-detection prospective, it is clear that when using \method to select users to attack, the time at which fraudulent transactions are detected is significantly delayed.
The mean time-to-detection of fraudulent transactions crafted using the predicted amount strategy is delayed from 3.55 to 5.81 (a 63\% improvement) and from 2.40 to 4.92 (a 105\% improvement) when using the pure random strategy.
With regard to research question \#2, we can conclude that \method is effective as a preliminary step of an attack.
The fraudable users found with our method turned out to be more vulnerable to fraud (with respect to fraud detection performance on them) than the users selected randomly.

\textbf{Target fraud detector}. Among the three tested target fraud detectors, OC-SVM, LOF and VAE, the latter showed the least decline in performance when applied on users chosen by fraudability scores.
Nevertheless, this outcome is somewhat obvious because, although having a different configuration, the VAE target model reassembles to some extent the surrogate fraud detector used to craft the attack.
In relation to the other two target models, the OC-SVM is slightly less robust to the attack, with an average 2\% increase in injection rate, when compared to LOF. 

\textbf{Injection strategy.} The most powerful injection strategies are the late detection and the profit maximization strategies, which take advantage of adversarial learning techniques to craft fraudulent transactions that maximize the time-to-detection and amount stolen, respectively.
Adversarial late detection achieved a detection time of 9.44 with logistic regression, while profit maximization was able to steal €3,348 -- both with the least amount of data available.
These two injection strategies, along with the predicted amount strategy, are more sensitive to a reduction in the amount of attacker knowledge, although this is to be expected, as these strategies include an attacker-trained model which is used to craft their malicious transactions.

\textbf{Required data.} Although a slight decrease in performance (a decrease of $\sim$5-10\% in the injection rate, $\sim$1 time unit in the time-to-detection, and €100 in the money stolen), reasonable results were achieved with only the accounts of just 320 users at the disposal of the attacker, which is only 5\% of the amount of data the defender possesses.
This means that the task of predicting the performance of a fraud detection system is not as difficult as other ML tasks, such as image classification, natural language processing, and drug discovery.
We assume that this is the case due to the human nature of the task of fraud susceptibility prediction where there are likely to be repeated patterns in the buying habits of some groups (or clusters) of users.

\textbf{Algorithm used.} It can be seen that the fraudability score predictor used on the random forest regression algorithm is superior in terms of the injection rate, time-to-detection, and money stolen.
This means that an attacker that implements the method proposed in this work is advised to use a random forest regression based algorithms in order to build the FRAUDability framework and predict the users' susceptibility to financial fraud.
In contrast, the linear regression algorithm achieved relatively poor performance, likely due to nonlinear relationships in the data.

\textbf{Importance of the results.} Clearly the findings in this study are valuable for an attacker that wishes to stay under the radar of AI-based fraud detectors. 
FRAUDability acts as a powerful compass, enabling the attacker to engage with only highly fraudable users, eliminating the effort and decreased financial gain of attacking less fraudable accounts.
However, the findings of this study are also valuable for the defending side, which pays the price of fraud in terms of financial loss and public image. 
The defender can establish a fraudability assessment environment (i.e., using our proposed \method method or a similar one) on site to identify highly fraudable users. 
With this knowledge, the defender can invest time and resources in providing extra protection to those accounts and refrain from protecting the less fraudable accounts which do not require extra fraud protection.
Such protection could consist of multi-factor authentication (MFA) or human expert analysis; alternatively, the defender could set lower thresholds for susceptible users.

\subsection{Effectiveness of the proposed adversarial attack}
Even without taking advantage of the fraudability scores, our proposed sequence-based adversarial attack demonstrated significant improvement over the two baselines -- the pure random and predicted amount strategies.
Having said that, it can be seen that although slight improvement in metrics when generating adversarial transaction using BIM instead of FGSM, this improvement is not significant.
From the injection rate perspective, the late detection strategy obtained an injection rate of 92.84\%, which is an increase of 157\% over the predicted amount strategy and an increase of 281\% over the pure random strategy.
Unsurprisingly, the late detection strategy also outperforms its counterparts with regard to the time-to-detection metric.
It introduced the latest-to-be-detected adversarial transactions with an average detection time of 9.21.
This average detection time is an improvement of 159\% over the predicted amount strategy and of 283\% over the pure random strategy.
From the amount of money stolen perspective, the profit maximization strategy was able to steal more money, stealing €2,747, while the pure random and predicted amount strategies respectively stole €215 and €244.
These results reflect the effectiveness of our proposed sequence-based adversarial attack against fraud detection systems.
With regard to research question \#3, we can conclude that our adversarial attack outperformed the traditional ways of injection fraud (i.e., randomizing or guessing).

\subsection{Countermeasures}
Our paper claims that attacker that is equipped with fraudability scores, can evade fraud detection with high probability.
Possible countermeasure might be applying multiple fraud detectors at the defending side (defense-in-depth).
In such a way, when one's fraud detector reliability is compromised by fraudability scores, other fraud detectors will increase the robustness of the system.
Another countermeasure to an attacker having fraudability scores calculated, is the defending side calculating his own set of fraudability scores beforehand.
Then, the defending side may harden the security of the fraudable users, by applying solution such as Multi-Factor Authentication (MFA).

\subsection{Explainability of the results}
Risk score methods are often required to be explainable in order to be trusted.
In order to explain our method's scoring, we further inspected the fraudability scores and the users' profiles to be able to explain the results.
We could see noticeable correlation between the 'avgAmount', 'avgTimeSinceLastTransaction' and 'varAmount' features (the average purchase amount, time gap between consequent transaction and the amount variance, respectively) and a high fraudability scores.
This could indicate a correlation between some user's fraudability score and the variance/jitteriness off its transactions.
This kind of inspection can provide some insights that will yield to security-related decisions in the organization.
Then, a new set of fraudability scores can be compiled, yielding to some new insights and so on, in an iterative manner.
\section{Limitations and Future Work} \label{limitations_future_work}

In addition to the assumptions made in Section \ref{sec:approach}, there are few more issues that need to be addressed.
As mentioned earlier, because of privacy and confidentiality issues, real financial data are scarce.
Although we had access to a real dataset provided by an online e-commerce platform, we still had to use unsupervised ML techniques to classify the manually injected fraudulent transactions to compensate for the lack of labels. 
However, to minimize the impact of this limitation, we accurately modeled injected fraudulent transactions in various ways, covering most of the ways a fraudster would operate. 
There is, however, a possible limitation regarding the source of the training and test sets in the experiments.
Although we built our attacker and defender's knowledge from disjoint sets of users, they both come from the same platform, in data collected during an approximately two-year period.
Therefore, we were unable to evaluate \method's ability to generalize \textbf{across datasets}, collected over a longer period of time, which would be an interesting direction for future work. 
Financial institutions have significantly more flexibility with training data, and can build robust models that can be retrain periodically.

Another interesting area to consider in future work involves exploring how fraudability scores could change in real-time, given the online nature of transactions today.
Given that it is possible for an attacker that has taken control of a compromised user account to start executing transactions on behalf of a victim, which themselves would change the fraudability score on the fly, it would be interesting to explore how the score could change and evaluate this property on a large scale.
This could perhaps be considered as moving towards a meta-score that relates to some characteristics (e.g., variability or stability) of the original fraudabiliy score presented in this work.
\section{Conclusion} \label{sec:conclusion}

In this paper, we presented \method, a novel approach for estimating the performance of a fraud detector for a given user.
We validated our approach by simulating an attacker that uses different attack strategies to perform an attack against state-of-the-art AI-based fraud detection systems.
We considered both the attacker and defender perspectives in this research.
Our approach assumes that a financial dataset is available to the attacker and that he/she can control a user's transactions.
From a defender standpoint (i.e., the service owner) this assumption is a given.

The results of our experiments show that a reasonable injection rate (i.e., the rate at which a transaction is not detected in all of the sequences the fraud detector considers in its classification) is achievable, even in the case of an attacker with the least amount of knowledge we evaluated -- when the attacker has only 5\% the defender's data.

The results show that using \method to generate fraudability scores can help the attacker identify users that are not well protected by the fraud detector, as the attack performed against fraudable users was significantly more effective.
In addition, our sequence-based adversarial attack was shown to generate the most effective adversarial transactions, with injection rates ranging from 89 to 98\%, the largest amounts stolen and latest detected transactions.

An interesting future challenge would be to both examine the ability of our method to generalize to different datasets and exploring how fraudability scores change over time, in an online manner.
The method proposed in this paper represents a significant contribution to the fraud detection domain, paving the way to more sophisticated ML-based attack and defense mindsets. On the one hand, it allows the defender to increase the robustness of his/her service by locating weak spots. On the other hand, it is capable of providing recommendations to an attacker regarding when, how, and at what scale to engage with a target victim.

\bibliographystyle{IEEEtran}
\bibliography{references}

\begin{thebibliography}{10}
\providecommand{\url}[1]{#1}
\csname url@samestyle\endcsname
\providecommand{\newblock}{\relax}
\providecommand{\bibinfo}[2]{#2}
\providecommand{\BIBentrySTDinterwordspacing}{\spaceskip=0pt\relax}
\providecommand{\BIBentryALTinterwordstretchfactor}{4}
\providecommand{\BIBentryALTinterwordspacing}{\spaceskip=\fontdimen2\font plus
\BIBentryALTinterwordstretchfactor\fontdimen3\font minus
  \fontdimen4\font\relax}
\providecommand{\BIBforeignlanguage}[2]{{%
\expandafter\ifx\csname l@#1\endcsname\relax
\typeout{** WARNING: IEEEtran.bst: No hyphenation pattern has been}%
\typeout{** loaded for the language `#1'. Using the pattern for}%
\typeout{** the default language instead.}%
\else
\language=\csname l@#1\endcsname
\fi
#2}}
\providecommand{\BIBdecl}{\relax}
\BIBdecl

\bibitem{mercer1990fraud}
L.~C. Mercer, ``Fraud detection via regression analysis,'' \emph{Computers \&
  Security}, vol.~9, no.~4, pp. 331--338, 1990.

\bibitem{22_malekian2013adaptive}
\BIBentryALTinterwordspacing
D.~Malekian and M.~R. Hashemi, ``An adaptive profile based fraud detection
  framework for handling concept drift,'' in \emph{10th International {ISC}
  Conference on Information Security and Cryptology, {ISCISC} 2013, Yazd, Iran,
  August 29-30, 2013}.\hskip 1em plus 0.5em minus 0.4em\relax {IEEE}, 2013, pp.
  1--6. [Online]. Available: \url{https://doi.org/10.1109/ISCISC.2013.6767338}
\BIBentrySTDinterwordspacing

\bibitem{jurgovsky2018sequence}
J.~Jurgovsky, M.~Granitzer, K.~Ziegler, S.~Calabretto, P.-E. Portier,
  L.~He-Guelton, and O.~Caelen, ``Sequence classification for credit-card fraud
  detection,'' \emph{Expert Systems with Applications}, vol. 100, pp. 234--245,
  2018.

\bibitem{DBLP:conf/raid/CarminatiSPZ20}
\BIBentryALTinterwordspacing
M.~Carminati, L.~Santini, M.~Polino, and S.~Zanero, ``Evasion attacks against
  banking fraud detection systems,'' in \emph{23rd International Symposium on
  Research in Attacks, Intrusions and Defenses, {RAID} 2020, San Sebastian,
  Spain, October 14-15, 2020}, M.~Egele and L.~Bilge, Eds.\hskip 1em plus 0.5em
  minus 0.4em\relax {USENIX} Association, 2020, pp. 285--300. [Online].
  Available:
  \url{https://www.usenix.org/conference/raid2020/presentation/carminati}
\BIBentrySTDinterwordspacing

\bibitem{DBLP:conf/www/GuoLAHHZZ19}
\BIBentryALTinterwordspacing
Q.~Guo, Z.~Li, B.~An, P.~Hui, J.~Huang, L.~Zhang, and M.~Zhao, ``Securing the
  deep fraud detector in large-scale e-commerce platform via adversarial
  machine learning approach,'' in \emph{The World Wide Web Conference, {WWW}
  2019, San Francisco, CA, USA, May 13-17, 2019}, L.~Liu, R.~W. White,
  A.~Mantrach, F.~Silvestri, J.~J. McAuley, R.~Baeza{-}Yates, and L.~Zia,
  Eds.\hskip 1em plus 0.5em minus 0.4em\relax {ACM}, 2019, pp. 616--626.
  [Online]. Available: \url{https://doi.org/10.1145/3308558.3313533}
\BIBentrySTDinterwordspacing

\bibitem{continella2017prometheus}
A.~Continella, M.~Carminati, M.~Polino, A.~Lanzi, S.~Zanero, and F.~Maggi,
  ``Prometheus: Analyzing webinject-based information stealers,'' \emph{Journal
  of Computer Security}, vol.~25, no.~2, pp. 117--137, 2017.

\bibitem{17_thenilsonreport}
D.~Robertson, ``The nilson report,'' in \emph{The Nilson Report, issue
  1164}.\hskip 1em plus 0.5em minus 0.4em\relax Santa Barbara, CA 93150, USA:
  The Nilson Report, 2019, pp. 24--29.

\bibitem{12_qi2018fintech}
Y.~Qi and J.~Xiao, ``Fintech: Ai powers financial services to improve people's
  lives,'' \emph{Communications of the ACM}, vol.~61, no.~11, pp. 65--69, 2018.

\bibitem{13_abdallah2016fraud}
A.~Abdallah, M.~A. Maarof, and A.~Zainal, ``Fraud detection system: A survey,''
  \emph{Journal of Network and Computer Applications}, vol.~68, pp. 90--113,
  2016.

\bibitem{10_mozaffari2014systematic}
M.~Mozaffari-Kermani, S.~Sur-Kolay, A.~Raghunathan, and N.~K. Jha, ``Systematic
  poisoning attacks on and defenses for machine learning in healthcare,''
  \emph{IEEE journal of biomedical and health informatics}, vol.~19, no.~6, pp.
  1893--1905, 2014.

\bibitem{15_randhawa2018credit}
K.~Randhawa, C.~K. Loo, M.~Seera, C.~P. Lim, and A.~K. Nandi, ``Credit card
  fraud detection using adaboost and majority voting,'' \emph{IEEE access},
  vol.~6, pp. 14\,277--14\,284, 2018.

\bibitem{14_bhattacharyya2011data}
S.~Bhattacharyya, S.~Jha, K.~Tharakunnel, and J.~C. Westland, ``Data mining for
  credit card fraud: A comparative study,'' \emph{Decision Support Systems},
  vol.~50, no.~3, pp. 602--613, 2011.

\bibitem{DBLP:conf/kdd/LiuZASLFHT20}
\BIBentryALTinterwordspacing
C.~Liu, Q.~Zhong, X.~Ao, L.~Sun, W.~Lin, J.~Feng, Q.~He, and J.~Tang, ``Fraud
  transactions detection via behavior tree with local intention calibration,''
  in \emph{{KDD} '20: The 26th {ACM} {SIGKDD} Conference on Knowledge Discovery
  and Data Mining, Virtual Event, CA, USA, August 23-27, 2020}, R.~Gupta,
  Y.~Liu, J.~Tang, and B.~A. Prakash, Eds.\hskip 1em plus 0.5em minus
  0.4em\relax {ACM}, 2020, pp. 3035--3043. [Online]. Available:
  \url{https://doi.org/10.1145/3394486.3403354}
\BIBentrySTDinterwordspacing

\bibitem{2_carlini2017towards}
\BIBentryALTinterwordspacing
N.~Carlini and D.~A. Wagner, ``Towards evaluating the robustness of neural
  networks,'' in \emph{2017 {IEEE} Symposium on Security and Privacy, {SP}
  2017, San Jose, CA, USA, May 22-26, 2017}.\hskip 1em plus 0.5em minus
  0.4em\relax {IEEE} Computer Society, 2017, pp. 39--57. [Online]. Available:
  \url{https://doi.org/10.1109/SP.2017.49}
\BIBentrySTDinterwordspacing

\bibitem{DBLP:journals/corr/GrossePM0M16}
\BIBentryALTinterwordspacing
K.~Grosse, N.~Papernot, P.~Manoharan, M.~Backes, and P.~D. McDaniel,
  ``Adversarial perturbations against deep neural networks for malware
  classification,'' \emph{CoRR}, vol. abs/1606.04435, 2016. [Online].
  Available: \url{http://arxiv.org/abs/1606.04435}
\BIBentrySTDinterwordspacing

\bibitem{DBLP:journals/corr/abs-1801-01944}
\BIBentryALTinterwordspacing
N.~Carlini and D.~A. Wagner, ``Audio adversarial examples: Targeted attacks on
  speech-to-text,'' \emph{CoRR}, vol. abs/1801.01944, 2018. [Online].
  Available: \url{http://arxiv.org/abs/1801.01944}
\BIBentrySTDinterwordspacing

\bibitem{5_nelson2008exploiting}
B.~Nelson, M.~Barreno, F.~J. Chi, A.~D. Joseph, B.~I. Rubinstein, U.~Saini,
  C.~A. Sutton, J.~D. Tygar, and K.~Xia, ``Exploiting machine learning to
  subvert your spam filter.'' \emph{LEET}, vol.~8, pp. 1--9, 2008.

\bibitem{DBLP:journals/corr/abs-1903-05157}
\BIBentryALTinterwordspacing
A.~Boloor, X.~He, C.~D. Gill, Y.~Vorobeychik, and X.~Zhang, ``Simple physical
  adversarial examples against end-to-end autonomous driving models,''
  \emph{CoRR}, vol. abs/1903.05157, 2019. [Online]. Available:
  \url{http://arxiv.org/abs/1903.05157}
\BIBentrySTDinterwordspacing

\bibitem{DBLP:conf/aaai/HuangSYH19}
\BIBentryALTinterwordspacing
X.~Huang, Q.~Song, F.~Yang, and X.~Hu, ``Large-scale heterogeneous feature
  embedding,'' in \emph{The Thirty-Third {AAAI} Conference on Artificial
  Intelligence, {AAAI} 2019, The Thirty-First Innovative Applications of
  Artificial Intelligence Conference, {IAAI} 2019, The Ninth {AAAI} Symposium
  on Educational Advances in Artificial Intelligence, {EAAI} 2019, Honolulu,
  Hawaii, USA, January 27 - February 1, 2019}.\hskip 1em plus 0.5em minus
  0.4em\relax {AAAI} Press, 2019, pp. 3878--3885. [Online]. Available:
  \url{https://doi.org/10.1609/aaai.v33i01.33013878}
\BIBentrySTDinterwordspacing

\bibitem{DBLP:journals/corr/GoodfellowSS14}
\BIBentryALTinterwordspacing
I.~J. Goodfellow, J.~Shlens, and C.~Szegedy, ``Explaining and harnessing
  adversarial examples,'' in \emph{3rd International Conference on Learning
  Representations, {ICLR} 2015, San Diego, CA, USA, May 7-9, 2015, Conference
  Track Proceedings}, Y.~Bengio and Y.~LeCun, Eds., 2015. [Online]. Available:
  \url{http://arxiv.org/abs/1412.6572}
\BIBentrySTDinterwordspacing

\bibitem{DBLP:conf/iclr/KurakinGB17a}
\BIBentryALTinterwordspacing
A.~Kurakin, I.~J. Goodfellow, and S.~Bengio, ``Adversarial examples in the
  physical world,'' in \emph{5th International Conference on Learning
  Representations, {ICLR} 2017, Toulon, France, April 24-26, 2017, Workshop
  Track Proceedings}.\hskip 1em plus 0.5em minus 0.4em\relax OpenReview.net,
  2017. [Online]. Available: \url{https://openreview.net/forum?id=HJGU3Rodl}
\BIBentrySTDinterwordspacing

\bibitem{biggio2018wild}
\BIBentryALTinterwordspacing
B.~Biggio and F.~Roli, ``Wild patterns: Ten years after the rise of adversarial
  machine learning,'' in \emph{Proceedings of the 2018 ACM SIGSAC Conference on
  Computer and Communications Security}, ser. CCS '18.\hskip 1em plus 0.5em
  minus 0.4em\relax New York, NY, USA: Association for Computing Machinery,
  2018, p. 2154–2156. [Online]. Available:
  \url{https://doi.org/10.1145/3243734.3264418}
\BIBentrySTDinterwordspacing

\bibitem{wilson2009analytical}
J.~H. Wilson, ``An analytical approach to detecting insurance fraud using
  logistic regression,'' \emph{Journal of Finance and accountancy}, vol.~1,
  p.~1, 2009.

\bibitem{sahin2011detecting}
Y.~Sahin and E.~Duman, ``Detecting credit card fraud by ann and logistic
  regression,'' in \emph{2011 International Symposium on Innovations in
  Intelligent Systems and Applications}.\hskip 1em plus 0.5em minus 0.4em\relax
  IEEE, 2011, pp. 315--319.

\bibitem{DBLP:conf/ann/1995}
\BIBentryALTinterwordspacing
P.~J. Braspenning, F.~Thuijsman, and A.~J. M.~M. Weijters, Eds.,
  \emph{Artificial Neural Networks: An Introduction to {ANN} Theory and
  Practice}, ser. Lecture Notes in Computer Science, vol. 931.\hskip 1em plus
  0.5em minus 0.4em\relax Springer, 1995. [Online]. Available:
  \url{https://doi.org/10.1007/BFb0027019}
\BIBentrySTDinterwordspacing

\bibitem{liaw2002classification}
A.~Liaw, M.~Wiener \emph{et~al.}, ``Classification and regression by
  randomforest.''

\bibitem{DBLP:conf/kdd/ChenG16}
\BIBentryALTinterwordspacing
T.~Chen and C.~Guestrin, ``Xgboost: {A} scalable tree boosting system,'' in
  \emph{Proceedings of the 22nd {ACM} {SIGKDD} International Conference on
  Knowledge Discovery and Data Mining, San Francisco, CA, USA, August 13-17,
  2016}, B.~Krishnapuram, M.~Shah, A.~J. Smola, C.~C. Aggarwal, D.~Shen, and
  R.~Rastogi, Eds.\hskip 1em plus 0.5em minus 0.4em\relax {ACM}, 2016, pp.
  785--794. [Online]. Available: \url{https://doi.org/10.1145/2939672.2939785}
\BIBentrySTDinterwordspacing

\end{thebibliography}

\appendices
\section{Running Times} \label{appndix:running_times}

The benefits of using FRAUDability (for defenders and attackers alike) can also be illustrated by considering the running time perspective.
Our experiments show that estimating users' financial susceptibility using the naive evaluation (i.e., by using the inject-and-test procedure, i.e., Algorithm ~\ref{algorithm:inject-and-test}) is a time-consuming task, while evaluation using FRAUDability is accomplished as quickly as using a ML model on a modern system.
This is true, since FRAUDability is assumed to learn the latent information regarding financial susceptibility instead of performing a naive evaluation.
From a feasibility point of view, a naive approach (i.e., applying inject-and-test like algorithms), would take an unreasonable amount of time to complete.
Table~\ref{table:running_times} presents a comparison of the estimated running times of the naive approach and the running times measured when using FRAUDability on three of the most popular e-commerce platforms.
As can be seen in the table, a naive estimation of how long it would take to run the profit maximization attack for the entire eBay user dataset is around 65.21 years.
Therefore, large e-commerce corporations with hundreds of millions of users, which are sometimes liable for the frauds committed using their services (like eBay, Amazon, and Taobao) will find FRAUDability-like approaches helpful for obtaining immediate insights regarding the susceptibility of their users, saving them a significant amount of time and money.

\begin{table}[h]
\centering
\caption{Running times for the different injection strategies used in this research, along with the estimated running times for large e-commerce platforms. Measurements were made using a NVIDIA GeForce 1080 GTX GPU. Approximations were calculated by multiplying the number of users in the Time per User.}
\begin{adjustbox}{max width=\textwidth}
\begin{tabular}{ccccc}
                                                                    & \begin{tabular}[c]{@{}c@{}}Pure\\ Random\end{tabular} & \begin{tabular}[c]{@{}c@{}}Predicted\\ Amount\end{tabular} & \begin{tabular}[c]{@{}c@{}}Adv. Late \\ Detection\end{tabular} & \begin{tabular}[c]{@{}c@{}}Adv. Max. \\ Profit\end{tabular} \\ \hline
Time/User                                                       & 3.95 (s)                                              & 7.29 (s)                                                   & 10.09 (s)                                                      & 11.30 (s)                                                  \\
\begin{tabular}[c]{@{}c@{}}Our dataset \\ (6,504 users)\end{tabular} & 7.14 (h)                                              & 13.17 (h)                                                  & 18.23 (h)                                                      & 20.41 (h)                                                  \\
\begin{tabular}[c]{@{}c@{}}eBay \\ (182M users)\end{tabular}        & 22.79 (y)                                             & 42.07 (y)                                                  & 58.23 (y)                                                      & 65.21 (y)                                                  \\
\begin{tabular}[c]{@{}c@{}}Amazon\\ (310M users)\end{tabular}       & 38.82 (y)                                             & 71.66 (y)                                                  & 99.18 (y)                                                      & 111.07 (y)                                                 \\
\begin{tabular}[c]{@{}c@{}}TaoBao\\ (710M users)\end{tabular}       & 88.93 (y)                                             & 164.12 (y)                                                 & 227.16 (y)                                                     & 254.41 (y)                                                 \\
\begin{tabular}[c]{@{}c@{}}FRAUDability\\ our dataset\end{tabular}  & 2.12 (s)                                              & 2.01 (s)                                                   & 2.45 (s)                                                       & 2.11 (s)                                                  
\end{tabular}
\end{adjustbox}
\label{table:running_times}
\end{table}

\section{Additional Target Models} \label{appndix:target_models}

In Section ~\ref{sec:experimental_evaluation} we presented the evaluation results of our method with OC-SVM as the target (attacked) fraud detector model.
To showcase the value of our work, we present here additional experiments that were conducted, with different target models.
The first additional experiment was conducted with a Local Outlier Factor (LOF) as the target model.
LOF is a popular unsupervided anomaly detection model for finding anomalous data points, by measuring the local deviation of a given data point with respect to its neighbours.
Our LOF-based target fraud detector was created with 20 neighbors and leaf size of 30, and the results are brought in Table~\ref{table:FRAUDability_results_2}.
It can be seen that when using LOF as the target model, an attacker can get similar achievements as before.
More specifically, an increase of 77\% (23.38\%→41.55\%) in injection rate in the case of a random injection strategy, and more than 3 times the amount of stolen money, when compared to the baseline.
Finally, Table~\ref{table:FRAUDability_results_3} present the results of a an experiment that was done with Variational Autoencoder (VAE) as the target model.
Yet, our tested VAE model was constructed using different architecture and parameters that the VAE that was used as the surrogate fraud detector.
Surprisingly enough, when comparing the results of this experiment to the those described in Table~\ref{table:FRAUDability_results_2}, the improvement over the baseline is a bit less prominent, with an increase of 67\% (26.76\%→44.84\%) in injection rate in the case of a random injection strategy, and less than 2 times the amount of stolen money.
We suspect that a possible explanation might be the better performance of VAE in detecting fraud in the baseline itself, leaving less room for the attack to improve.

\begin{table*}[ht]
\footnotesize
\centering
\caption{Experimental results for all injection strategies, attacker knowledge levels, and FRAUDability implementations, on a LOF target (attacked) fraud detector. The best results from the attacker's points of view appear in {\color[HTML]{009901} green}, and the worst results from the attacker's point of view appear in \textcolor{red}{red}. The values are averaged over the number of users evaluated. Adversarial Max Profit and Adversarial Late Detection results are formatted \textless FGSM\textgreater |\textless BIM\textgreater.
}
\resizebox{\textwidth}{!}
{
\begin{tabular}{ccccccc}
\multicolumn{1}{l}{}                                                                                                 &                              & \multicolumn{4}{c}{FRAUDability}                                                                                                            & Random Selection     \\ \hline
\multicolumn{1}{c|}{}                                                                                                & Metric/Model                 & Neural Network                 & RandomForest Regressor                   & XGBoost Regressor              & Linear Regression              & \multicolumn{1}{l}{} \\ \hline
\multicolumn{7}{c}{Strategy: Pure Random}                                                                                                                                                                                                                                                                                \\ \hline
\multicolumn{1}{c|}{}                                                                                                & Injection Rate               & 39.85\%                        & {\color[HTML]{009901} 41.55\%}           & 35.66\%                        & {\color[HTML]{FE0000} 32.28\%} & 23.38\%              \\
\multicolumn{1}{c|}{}                                                                                                & Time-to-Detection {[}0-10{]} & 5.33                           & {\color[HTML]{009901} 5.62}              & 5.21                           & {\color[HTML]{FE0000} 4.38}    & 2.50                 \\
\multicolumn{1}{c|}{\multirow{-3}{*}{\begin{tabular}[c]{@{}c@{}}Defender's \\ Knowledge:\\ 6400 users\end{tabular}}} & Money Stolen                 & € 764                          & {\color[HTML]{009901} € 773}             & € 624                          & {\color[HTML]{FE0000} € 621}   & € 212.46             \\ \hline
\multicolumn{1}{c|}{}                                                                                                & Injection Rate               & 36.08\%                        & {\color[HTML]{009901} 37.67\%}           & {\color[HTML]{000000} 33.68\%} & {\color[HTML]{FE0000} 30.85\%} & 23.38\%              \\
\multicolumn{1}{c|}{}                                                                                                & Time-to-Detection {[}0-10{]} & 4.37                           & {\color[HTML]{009901} 4.82}              & 4.13                           & {\color[HTML]{FE0000} 3.67}    & 2.50                 \\
\multicolumn{1}{c|}{\multirow{-3}{*}{\begin{tabular}[c]{@{}c@{}}Attacker's \\ Knowledge:\\ 1921 users\end{tabular}}} & Money Stolen                 & € 554                          & {\color[HTML]{009901} € 562}             & {\color[HTML]{000000} € 539}   & {\color[HTML]{FE0000} € 531}   & € 212.46             \\ \hline
\multicolumn{1}{c|}{}                                                                                                & Injection Rate               & {\color[HTML]{009901} 33.53\%} & {\color[HTML]{000000} 31.76\%}           & {\color[HTML]{000000} 32.66\%} & {\color[HTML]{FE0000} 29.98\%} & 23.38\%              \\
\multicolumn{1}{c|}{}                                                                                                & Time-to-Detection {[}0-10{]} & 3.94                           & {\color[HTML]{009901} 4.06}              & 3.16                           & {\color[HTML]{FE0000} 3.06}    & 2.50                 \\
\multicolumn{1}{c|}{\multirow{-3}{*}{\begin{tabular}[c]{@{}c@{}}Attacker's\\ Knowledge: \\ 1281 users\end{tabular}}} & Money Stolen                 & {\color[HTML]{009901} € 532}   & {\color[HTML]{FE0000} € 506}             & € 523                          & € 530                          & € 212.46             \\ \hline
\multicolumn{1}{c|}{}                                                                                                & Injection Rate               & {\color[HTML]{009901} 31.82\%} & 28.9\%                                   & 29.71\%                        & {\color[HTML]{FE0000} 28.37\%} & 23.38\%              \\
\multicolumn{1}{c|}{}                                                                                                & Time-to-Detection {[}0-10{]} & {\color[HTML]{009901} 3.65}    & 3.57                                     & 2.39                           & {\color[HTML]{FE0000} 2.32}    & 2.50                 \\
\multicolumn{1}{c|}{\multirow{-3}{*}{\begin{tabular}[c]{@{}c@{}}Attacker's \\ Knowledge: \\ 320 users\end{tabular}}} & Money Stolen                 & {\color[HTML]{009901} € 530}   & € 444                                    & {\color[HTML]{FE0000} € 419}   & {\color[HTML]{333333} € 457}   & € 212.46             \\ \hline
\multicolumn{7}{c}{Strategy: Predicted Amount}                                                                                                                                                                                                                                                                           \\ \hline
\multicolumn{1}{c|}{}                                                                                                & Injection Rate               & 61.65\%                        & {\color[HTML]{009901} 66.75\%}           & {\color[HTML]{FE0000} 53.53\%} & 64.23\%                        & 33.32\%              \\
\multicolumn{1}{c|}{}                                                                                                & Time-to-Detection {[}0-10{]} & 6.87                           & {\color[HTML]{009901} 7.32}              & 6.32                           & {\color[HTML]{FE0000} 5.86}    & 3.43                 \\
\multicolumn{1}{c|}{\multirow{-3}{*}{\begin{tabular}[c]{@{}c@{}}Defender's \\ Knowledge:\\ 6400 users\end{tabular}}} & Money Stolen                 & € 632                          & {\color[HTML]{009901} € 776}             & {\color[HTML]{333333} € 548}   & {\color[HTML]{FE0000} € 530}   & € 224.42             \\ \hline
\multicolumn{1}{c|}{}                                                                                                & Injection Rate               & 55.61\%                        & {\color[HTML]{009901} 57.05\%}           & {\color[HTML]{FE0000} 48.24\%} & 54.43\%                        & 33.32\%              \\
\multicolumn{1}{c|}{}                                                                                                & Time-to-Detection {[}0-10{]} & 5.55                           & {\color[HTML]{009901} 5.69}              & 4.94                           & {\color[HTML]{FE0000} 4.67}    & 3.43                 \\
\multicolumn{1}{c|}{\multirow{-3}{*}{\begin{tabular}[c]{@{}c@{}}Attacker's \\ Knowledge:\\ 1921 users\end{tabular}}} & Money Stolen                 & € 436                          & € 529                                    & {\color[HTML]{009901} € 531}   & {\color[HTML]{FE0000} € 440}   & € 224.42             \\ \hline
\multicolumn{1}{c|}{}                                                                                                & Injection Rate               & 53.04\%                        & {\color[HTML]{009901} 56.89\%}           & 51.18\%                        & {\color[HTML]{FE0000} 48.24\%} & 33.32\%              \\
\multicolumn{1}{c|}{}                                                                                                & Time-to-Detection {[}0-10{]} & 5.31                           & {\color[HTML]{009901} 5.33}              & 4.74                           & {\color[HTML]{FE0000} 4.38}    & 3.43                 \\
\multicolumn{1}{c|}{\multirow{-3}{*}{\begin{tabular}[c]{@{}c@{}}Attacker's\\ Knowledge: \\ 1281 users\end{tabular}}} & Money Stolen                 & € 511                          & {\color[HTML]{009901} € 535}             & € 521                          & {\color[HTML]{FE0000} € 401}   & € 224.42             \\ \hline
\multicolumn{1}{c|}{}                                                                                                & Injection Rate               & 49.2\%                         & {\color[HTML]{009901} 54.4\%}            & 49.4\%                         & {\color[HTML]{FE0000} 39.44\%} & 33.32\%              \\
\multicolumn{1}{c|}{}                                                                                                & Time-to-Detection {[}0-10{]} & 5.11                           & {\color[HTML]{009901} 5.14}              & 4.13                           & {\color[HTML]{FE0000} 3.16}    & 3.43                 \\
\multicolumn{1}{c|}{\multirow{-3}{*}{\begin{tabular}[c]{@{}c@{}}Attacker's \\ Knowledge: \\ 320 users\end{tabular}}} & Money Stolen                 & € 489                          & {\color[HTML]{009901} € 499}             & €457                           & {\color[HTML]{FE0000} € 395}   & € 224.42             \\ \hline
\multicolumn{7}{c}{Strategy: Adversarial Late Detection}                                                                                                                                                                                                                                                                                                   \\ \hline
\multicolumn{1}{c|}{}                                                                                                & Injection Rate               & 95.81\% | 95.77\%                         & {\color[HTML]{009901} 96.74\% | 96.27\%}  & 95.1\% | 95.66 \%                        & {\color[HTML]{FE0000} 92.38\% | 92.42 \%} & 92.84\%              \\
\multicolumn{1}{c|}{}                                                                                                & Time-to-Detection {[}0-10{]} & 9.68 | 9.69                               & {\color[HTML]{009901} 9.69 | 9.68}        & {\color[HTML]{330001} 9.42 | 9.49}        & {\color[HTML]{FE0000} 9.03 | 9.11}        & 9.21                 \\
\multicolumn{1}{c|}{\multirow{-3}{*}{\begin{tabular}[c]{@{}c@{}}Defender's \\ Knowledge:\\ 6400 users\end{tabular}}} & Money Stolen                 & € 3204 | € 3300                           & {\color[HTML]{009901} € 3343 | € 3303}    & € 3199 | € 3221                           & {\color[HTML]{FE0000} € 3014 | € 3015}    & € 2532.46            \\ \hline
\multicolumn{1}{c|}{}                                                                                                & Injection Rate               & 96.01\% | 96.88 \%                        & {\color[HTML]{009901} 97.42\% | 98.73 \%} & 95.88\% | 96.45 \%                        & {\color[HTML]{FE0000} 91.18\% | 91.44 \%} & 92.84\%              \\
\multicolumn{1}{c|}{}                                                                                                & Time-to-Detection {[}0-10{]} & 9.41 | 9.70                               & {\color[HTML]{009901} 9.42 | 9.70}        & {\color[HTML]{330001} 9.49 | 9.58}        & {\color[HTML]{FE0000} 9.21 | 9.29}        & 9.21                 \\
\multicolumn{1}{c|}{\multirow{-3}{*}{\begin{tabular}[c]{@{}c@{}}Attacker's \\ Knowledge:\\ 1921 users\end{tabular}}} & Money Stolen                 & € 3076 | € 3088                           & {\color[HTML]{009901} € 3100 | € 3190}    & € 3026 | € 3046                           & {\color[HTML]{FE0000} € 2932 | € 3020}    & € 2532.46            \\ \hline
\multicolumn{1}{c|}{}                                                                                                & Injection Rate               & 93.65\% | 94.19 \%                        & {\color[HTML]{009901} 95.18\% | 96.06 \%} & 93.12\% | 94.09 \%                        & {\color[HTML]{FE0000} 88.21\% | 88.31 \%} & 92.84\%              \\
\multicolumn{1}{c|}{}                                                                                                & Time-to-Detection {[}0-10{]} & {\color[HTML]{009901} 9.55 | 9.74}        & {\color[HTML]{000000} 9.36 | 9.43}        & 9.26 | 9.32                               & {\color[HTML]{FE0000} 9.21 | 9.26}        & 9.21                 \\
\multicolumn{1}{c|}{\multirow{-3}{*}{\begin{tabular}[c]{@{}c@{}}Attacker's\\ Knowledge: \\ 1281 users\end{tabular}}} & Money Stolen                 & € 2821 | € 2852                           & {\color[HTML]{009901} € 2910 | € 2949}    & € 2841 | € 2845                           & {\color[HTML]{FE0000} € 2766 | € 2786}    & € 2532.46            \\ \hline
\multicolumn{1}{c|}{}                                                                                                & Injection Rate               & 92.36\% | 93.14 \%                        & {\color[HTML]{009901} 93.83\% | 94.19 \%} & 91.34\% | 92.29 \%                         & {\color[HTML]{FE0000} 87.98\% | 88.31 \%} & 92.84\%              \\
\multicolumn{1}{c|}{}                                                                                                & Time-to-Detection {[}0-10{]} & 9.11 | 9.26                               & {\color[HTML]{FE0000} 9.1 | 9.12}        & 9.22 | 9.31                               & {\color[HTML]{009901} 9.25 | 9.32}        & 9.21                 \\
\multicolumn{1}{c|}{\multirow{-3}{*}{\begin{tabular}[c]{@{}c@{}}Attacker's \\ Knowledge: \\ 320 users\end{tabular}}} & Money Stolen                 & € 2753 | € 2827                           & {\color[HTML]{009901} € 2817 | € 2831}    & {\color[HTML]{FE0000} € 2492 | € 2589}    & € 2623 | € 2629                           & € 2532.46            \\ \hline
\multicolumn{7}{c}{Strategy: Adversarial Max Profit}                                                                                                                                                                                                                                                                                                       \\ \hline
                                                                                                                     & Injection Rate               & 92.38\% | 93.02 \%                        & 93.36\% | 94.11 \%                        & {\color[HTML]{009901} 94.88\% | 95.17 \%} & {\color[HTML]{FE0000} 90.54\% | 91.23 \%} & 87.44\%              \\
                                                                                                                     & Time-to-Detection {[}0-10{]} & 8.30 | 8.40                               & 8.40 | 8.60                               & {\color[HTML]{009901} 9.00 | 9.10}        & {\color[HTML]{FE0000} 8.20 | 8.40}        & 7.18                 \\
\multirow{-3}{*}{\begin{tabular}[c]{@{}c@{}}Defender's \\ Knowledge:\\ 6400 users\end{tabular}}                      & Money Stolen                 & € 3923 | € 3994                           & {\color[HTML]{FE0000} € 3600 | € 3623}    & {\color[HTML]{009901} € 4127 | € 4169}    & € 3891 | € 3920                           & € 2692            \\ \hline
\multicolumn{1}{c|}{}                                                                                                & Injection Rate               & 89.44\% | 89.5 \%                        & 88.66\% | 89.71 \%                        & {\color[HTML]{009901} 89.65\% | 89.83 \%} & {\color[HTML]{FE0000} 83.96\% | 84.18 \%} & 87.44\%              \\
\multicolumn{1}{c|}{}                                                                                                & Time-to-Detection {[}0-10{]} & 7.01 | 7.20                               & 7.20 | 7.19                               & {\color[HTML]{009901} 7.20 | 7.30}        & {\color[HTML]{FE0000} 7.00 | 7.11}        & 7.18                 \\
\multicolumn{1}{c|}{\multirow{-3}{*}{\begin{tabular}[c]{@{}c@{}}Attacker's \\ Knowledge:\\ 1921 users\end{tabular}}} & Money Stolen                 & € 3431 | € 3490                           & {\color[HTML]{FE0000} € 3387 | € 3440}    & {\color[HTML]{009901} € 3509 | € 3529}    & € 3433 | € 3493                           & € 2747.48            \\ \hline
\multicolumn{1}{c|}{}                                                                                                & Injection Rate               & {\color[HTML]{009901} 88.91\% | 90.7 \%} & 87.72\% | 90.00 \%                        & 89.01\% | 89.56 \%                        & {\color[HTML]{FE0000} 82.08\% | 82.87 \%} & 87.44\%              \\
\multicolumn{1}{c|}{}                                                                                                & Time-to-Detection {[}0-10{]} & {\color[HTML]{FE0000} 7.01 | 7.15}        & {\color[HTML]{000000} 6.96 | 7.1}        & {\color[HTML]{FE0000} 7.83 | 8.0}        & 6.88 | 6.89                               & 7.18                 \\
\multicolumn{1}{c|}{\multirow{-3}{*}{\begin{tabular}[c]{@{}c@{}}Attacker's\\ Knowledge: \\ 1281 users\end{tabular}}} & Money Stolen                 & {\color[HTML]{000000} € 3375 | € 3491}    & € 3331 | € 3383                           & {\color[HTML]{009901} € 3412 | € 3440}    & {\color[HTML]{FE0000} € 2937 | € 2957}    & € 2747.48            \\ \hline
\multicolumn{1}{c|}{}                                                                                                & Injection Rate               & 86.95\% | 87.54 \%                        & {\color[HTML]{009901} 87.62\% | 88.5 \%} & 87.36\% | 88.61 \%                        & {\color[HTML]{FE0000} 81.33\% | 82.54 \%} & 87.44\%              \\
\multicolumn{1}{c|}{}                                                                                                & Time-to-Detection {[}0-10{]} & {\color[HTML]{009901} 7.83 | 8.15}        & {\color[HTML]{000000} 6.91 | 6.98}        & 7.71 | 7.85                               & {\color[HTML]{FE0000} 6.89 | 7.28}        & 7.18                 \\
\multicolumn{1}{c|}{\multirow{-3}{*}{\begin{tabular}[c]{@{}c@{}}Attacker's \\ Knowledge: \\ 320 users\end{tabular}}} & Money Stolen                 & € 3233 | € 3247                           & {\color[HTML]{009901} € 3281 | € 3396}    & € 3268 | € 3348                           & {\color[HTML]{FE0000} € 3175 | € 3266}    & € 2747.48            \\ \hline
\end{tabular}
}
\label{table:FRAUDability_results_2}
\end{table*}

\begin{table*}[ht]
\footnotesize
\centering
\caption{Experimental results for all injection strategies, attacker knowledge levels, and FRAUDability implementations, on a VAE target (attacked) fraud detector. The best results from the attacker's points of view appear in {\color[HTML]{009901} green}, and the worst results from the attacker's point of view appear in \textcolor{red}{red}. The values are averaged over the number of users evaluated. Adversarial Max Profit and Adversarial Late Detection results are formatted as FGSM | BIM.
}
\resizebox{\textwidth}{!}
{
\begin{tabular}{ccccccc}
\multicolumn{1}{l}{}                                                                                                 &                              & \multicolumn{4}{c}{FRAUDability}                                                                                                                                              & Random Selection     \\ \hline
\multicolumn{1}{c|}{}                                                                                                & Metric/Model                 & Neural Network                            & RandomForest Regressor                    & XGBoost Regressor                         & Linear Regression                         & \multicolumn{1}{l}{} \\ \hline
\multicolumn{7}{c}{Strategy: Pure Random}                                                                                                                                                                                                                                                                                                                  \\ \hline
\multicolumn{1}{c|}{}                                                                                                & Injection Rate               & 42.14\%                                   & {\color[HTML]{009901} 44.84\%}            & 38.73\%                                   & {\color[HTML]{FE0000} 33.62\%}            & 26.76\%              \\
\multicolumn{1}{c|}{}                                                                                                & Time-to-Detection {[}0-10{]} & 6.46                                      & {\color[HTML]{009901} 6.74}               & 6.32                                      & {\color[HTML]{FE0000} 5.47}               & 3.40                 \\
\multicolumn{1}{c|}{\multirow{-3}{*}{\begin{tabular}[c]{@{}c@{}}Defender's \\ Knowledge:\\ 6400 users\end{tabular}}} & Money Stolen                 & € 780                                     & {\color[HTML]{009901} € 789}              & € 637                                     & {\color[HTML]{FE0000} € 700}              & € 215.37             \\ \hline
\multicolumn{1}{c|}{}                                                                                                & Injection Rate               & 38.82\%                                   & {\color[HTML]{009901} 38.44\%}            & {\color[HTML]{000000} 37.37\%}            & {\color[HTML]{FE0000} 31.48\%}            & 26.76\%              \\
\multicolumn{1}{c|}{}                                                                                                & Time-to-Detection {[}0-10{]} & 4.46                                      & {\color[HTML]{009901} 5.92}               & 5.21                                      & {\color[HTML]{FE0000} 3.74}               & 3.40                 \\
\multicolumn{1}{c|}{\multirow{-3}{*}{\begin{tabular}[c]{@{}c@{}}Attacker's \\ Knowledge:\\ 1921 users\end{tabular}}} & Money Stolen                 & € 566                                     & {\color[HTML]{009901} € 574}              & {\color[HTML]{000000} € 551}              & {\color[HTML]{FE0000} € 542}              & € 215.37             \\ \hline
\multicolumn{1}{c|}{}                                                                                                & Injection Rate               & {\color[HTML]{009901} 34.21\%}            & {\color[HTML]{000000} 32.41\%}            & {\color[HTML]{000000} 33.33 \%}           & {\color[HTML]{FE0000} 30.59\%}            & 26.76\%              \\
\multicolumn{1}{c|}{}                                                                                                & Time-to-Detection {[}0-10{]} & 4.02                                      & {\color[HTML]{009901} 4.14}               & 3.22                                      & {\color[HTML]{FE0000} 3.12}               & 2.40                 \\
\multicolumn{1}{c|}{\multirow{-3}{*}{\begin{tabular}[c]{@{}c@{}}Attacker's\\ Knowledge: \\ 1281 users\end{tabular}}} & Money Stolen                 & {\color[HTML]{009901} € 543}              & {\color[HTML]{FE0000} € 517}              & € 534                                     & € 541                                     & € 215.37             \\ \hline
\multicolumn{1}{c|}{}                                                                                                & Injection Rate               & {\color[HTML]{009901} 32.47\%}            & 29.49\%                                   & 30.32\%                                   & {\color[HTML]{FE0000} 28.95\%}            & 26.76\%              \\
\multicolumn{1}{c|}{}                                                                                                & Time-to-Detection {[}0-10{]} & {\color[HTML]{009901} 3.72}               & 3.64                                      & 2.44                                      & {\color[HTML]{FE0000} 2.37}               & 2.40                 \\
\multicolumn{1}{c|}{\multirow{-3}{*}{\begin{tabular}[c]{@{}c@{}}Attacker's \\ Knowledge: \\ 320 users\end{tabular}}} & Money Stolen                 & {\color[HTML]{009901} € 541}              & € 454                                     & {\color[HTML]{FE0000} € 428}              & {\color[HTML]{333333} € 467}              & € 215.37             \\ \hline
\multicolumn{7}{c}{Strategy: Predicted Amount}                                                                                                                                                                                                                                                                                                             \\ \hline
\multicolumn{1}{c|}{}                                                                                                & Injection Rate               & 62.91\%                                   & {\color[HTML]{009901} 72.09\%}            & {\color[HTML]{FE0000} 54.62\%}            & 65.54\%                                   & 39.6\%              \\
\multicolumn{1}{c|}{}                                                                                                & Time-to-Detection {[}0-10{]} & 7.01                                      & {\color[HTML]{009901} 7.47}               & 6.45                                      & {\color[HTML]{FE0000} 5.98}               & 4.55                 \\
\multicolumn{1}{c|}{\multirow{-3}{*}{\begin{tabular}[c]{@{}c@{}}Defender's \\ Knowledge:\\ 6400 users\end{tabular}}} & Money Stolen                 & € 645                                     & {\color[HTML]{009901} € 792}              & {\color[HTML]{333333} € 560}              & {\color[HTML]{FE0000} € 541}              & € 244.67             \\ \hline
\multicolumn{1}{c|}{}                                                                                                & Injection Rate               & 56.74\%                                   & {\color[HTML]{009901} 58.21\%}            & {\color[HTML]{FE0000} 49.22\%}            & 55.54\%                                   & 39.6\%              \\
\multicolumn{1}{c|}{}                                                                                                & Time-to-Detection {[}0-10{]} & 5.66                                      & {\color[HTML]{009901} 5.81}               & 5.04                                      & {\color[HTML]{FE0000} 4.77}               & 3.99                 \\
\multicolumn{1}{c|}{\multirow{-3}{*}{\begin{tabular}[c]{@{}c@{}}Attacker's \\ Knowledge:\\ 1921 users\end{tabular}}} & Money Stolen                 & € 445                                     & € 540                                     & {\color[HTML]{009901} € 542}              & {\color[HTML]{FE0000} € 449}              & € 244.67             \\ \hline
\multicolumn{1}{c|}{}                                                                                                & Injection Rate               & 55.14\%                                   & {\color[HTML]{009901} 58.05\%}            & 52.22 \%                                  & {\color[HTML]{FE0000} 49.22\%}            & 39.6\%              \\
\multicolumn{1}{c|}{}                                                                                                & Time-to-Detection {[}0-10{]} & 6.24                                      & {\color[HTML]{009901} 5.44}               & 5.48                                      & {\color[HTML]{FE0000} 4.47}               & 4.66                 \\
\multicolumn{1}{c|}{\multirow{-3}{*}{\begin{tabular}[c]{@{}c@{}}Attacker's\\ Knowledge: \\ 1281 users\end{tabular}}} & Money Stolen                 & € 522                                     & {\color[HTML]{009901} € 556}              & € 564                                     & {\color[HTML]{FE0000} € 510}              & € 279.67             \\ \hline
\multicolumn{1}{c|}{}                                                                                                & Injection Rate               & 52.65\%                                   & {\color[HTML]{009901} 58.51\%}            & 50.41 \%                                  & {\color[HTML]{FE0000} 46.42\%}            & 39.6\%              \\
\multicolumn{1}{c|}{}                                                                                                & Time-to-Detection {[}0-10{]} & 5.21                                      & {\color[HTML]{009901} 5.24}               & 4.21                                      & {\color[HTML]{FE0000} 3.22}               & 3.55                 \\
\multicolumn{1}{c|}{\multirow{-3}{*}{\begin{tabular}[c]{@{}c@{}}Attacker's \\ Knowledge: \\ 320 users\end{tabular}}} & Money Stolen                 & € 499                                     & {\color[HTML]{009901} € 510}              & € 467                                     & {\color[HTML]{FE0000} € 404}              & € 244.67             \\ \hline
\multicolumn{7}{c}{Strategy: Adversarial Late Detection}                                                                                                                                                                                                                                                                                                   \\ \hline
\multicolumn{1}{c|}{}                                                                                                & Injection Rate               & 97.77\% | 97.72\%                         & {\color[HTML]{009901} 98.71\% | 98.23\%}  & 97.04\% | 97.61 \%                        & {\color[HTML]{FE0000} 94.27\% | 94.31 \%} & 92.84\%              \\
\multicolumn{1}{c|}{}                                                                                                & Time-to-Detection {[}0-10{]} & 9.88 | 9.89                               & {\color[HTML]{009901} 9.89 | 9.88}        & {\color[HTML]{330001} 9.61 | 9.68}        & {\color[HTML]{FE0000} 9.21 | 9.30}        & 9.21                 \\
\multicolumn{1}{c|}{\multirow{-3}{*}{\begin{tabular}[c]{@{}c@{}}Defender's \\ Knowledge:\\ 6400 users\end{tabular}}} & Money Stolen                 & € 3304 | € 3309                           & {\color[HTML]{009901} € 3434 | € 3430}    & € 3231 | € 3233                           & {\color[HTML]{FE0000} € 3011 | € 3016}    & € 2532.46            \\ \hline
\multicolumn{1}{c|}{}                                                                                                & Injection Rate               & 96.67\% | 96.99 \%                        & {\color[HTML]{009901} 98.24\% | 98.91 \%} & 96.12\% | 96.63 \%                        & {\color[HTML]{FE0000} 91.14\% | 91.74 \%} & 92.84\%              \\
\multicolumn{1}{c|}{}                                                                                                & Time-to-Detection {[}0-10{]} & 9.74 | 9.74                               & {\color[HTML]{009901} 9.77 | 9.78}        & {\color[HTML]{330001} 9.54 | 9.69}        & {\color[HTML]{FE0000} 9.26 | 9.32}        & 9.21                 \\
\multicolumn{1}{c|}{\multirow{-3}{*}{\begin{tabular}[c]{@{}c@{}}Attacker's \\ Knowledge:\\ 1921 users\end{tabular}}} & Money Stolen                 & € 3087 | € 3095                           & {\color[HTML]{009901} € 3102 | € 3196}    & € 3033 | € 3047                           & {\color[HTML]{FE0000} € 2932 | € 3023}    & € 2532.46            \\ \hline
\multicolumn{1}{c|}{}                                                                                                & Injection Rate               & 95.56\% | 96.11 \%                        & {\color[HTML]{009901} 97.12\% | 98.02 \%} & 95.02\% | 96.01 \%                        & {\color[HTML]{FE0000} 90.01\% | 90.11 \%} & 92.84\%              \\
\multicolumn{1}{c|}{}                                                                                                & Time-to-Detection {[}0-10{]} & {\color[HTML]{009901} 9.74 | 9.94}        & {\color[HTML]{000000} 9.55 | 9.62}        & 9.45 | 9.51                               & {\color[HTML]{FE0000} 9.40 | 9.45}        & 9.21                 \\
\multicolumn{1}{c|}{\multirow{-3}{*}{\begin{tabular}[c]{@{}c@{}}Attacker's\\ Knowledge: \\ 1281 users\end{tabular}}} & Money Stolen                 & € 2879 | € 2911                           & {\color[HTML]{009901} € 2970 | € 3010}    & € 2899 | € 2904                           & {\color[HTML]{FE0000} € 2823 | € 2843}    & € 2532.46            \\ \hline
\multicolumn{1}{c|}{}                                                                                                & Injection Rate               & 94.24\% | 95.04 \%                        & {\color[HTML]{009901} 95.74\% | 96.11 \%} & 93.2\% | 94.17 \%                         & {\color[HTML]{FE0000} 89.78\% | 90.11 \%} & 92.84\%              \\
\multicolumn{1}{c|}{}                                                                                                & Time-to-Detection {[}0-10{]} & 9.76 | 9.45                               & {\color[HTML]{FE0000} 9.35 | 9.46}        & 9.41 | 9.50                               & {\color[HTML]{009901} 9.44 | 9.51}        & 9.21                 \\
\multicolumn{1}{c|}{\multirow{-3}{*}{\begin{tabular}[c]{@{}c@{}}Attacker's \\ Knowledge: \\ 320 users\end{tabular}}} & Money Stolen                 & € 2810 | € 2885                           & {\color[HTML]{009901} € 2875 | € 2889}    & {\color[HTML]{FE0000} € 2543 | € 2642}    & € 2677 | € 2683                           & € 2532.46            \\ \hline
\multicolumn{7}{c}{Strategy: Adversarial Max Profit}                                                                                                                                                                                                                                                                                                       \\ \hline
                                                                                                                     & Injection Rate               & 96.27\% | 97.92 \%                        & 95.27\% | 96.03 \%                        & {\color[HTML]{009901} 96.82\% | 98.14 \%} & {\color[HTML]{FE0000} 92.39\% | 93.09 \%} & 87.44\%              \\
                                                                                                                     & Time-to-Detection {[}0-10{]} & 8.37 | 8.46                               & 8.41 | 8.67                               & {\color[HTML]{009901} 9.03 | 9.13}        & {\color[HTML]{FE0000} 8.28 | 8.41}        & 7.18                 \\
\multirow{-3}{*}{\begin{tabular}[c]{@{}c@{}}Defender's \\ Knowledge:\\ 6400 users\end{tabular}}                      & Money Stolen                 & € 4004 | € 4076                           & {\color[HTML]{FE0000} € 3674 | € 3697}    & {\color[HTML]{009901} € 4212 | € 4255}    & € 3971 | € 4001                           & € 2747.48            \\ \hline
\multicolumn{1}{c|}{}                                                                                                & Injection Rate               & 91.27\% | 91.33 \%                        & 90.47\% | 91.54 \%                        & {\color[HTML]{009901} 91.48\% | 91.66 \%} & {\color[HTML]{FE0000} 85.67\% | 85.90 \%} & 87.44\%              \\
\multicolumn{1}{c|}{}                                                                                                & Time-to-Detection {[}0-10{]} & 7.15 | 7.24                               & 7.30 | 7.29                               & {\color[HTML]{009901} 7.31 | 7.40}        & {\color[HTML]{FE0000} 7.20 | 7.21}        & 7.18                 \\
\multicolumn{1}{c|}{\multirow{-3}{*}{\begin{tabular}[c]{@{}c@{}}Attacker's \\ Knowledge:\\ 1921 users\end{tabular}}} & Money Stolen                 & € 3502 | € 3562                           & {\color[HTML]{FE0000} € 3457 | € 3511}    & {\color[HTML]{009901} € 3581 | € 3602}    & € 3504 | € 3565                           & € 2747.48            \\ \hline
\multicolumn{1}{c|}{}                                                                                                & Injection Rate               & {\color[HTML]{009901} 90.72\% | 92.55 \%} & 89.51\% | 90.14 \%                        & 89.14\% | 90.11 \%                        & {\color[HTML]{FE0000} 83.75\% | 84.56 \%} & 87.44\%              \\
\multicolumn{1}{c|}{}                                                                                                & Time-to-Detection {[}0-10{]} & {\color[HTML]{FE0000} 7.01 | 7.15}        & {\color[HTML]{000000} 7.10 | 7.25}        & {\color[HTML]{FE0000} 7.99 | 8.16}        & 7.02 | 7.03                               & 7.18                 \\
\multicolumn{1}{c|}{\multirow{-3}{*}{\begin{tabular}[c]{@{}c@{}}Attacker's\\ Knowledge: \\ 1281 users\end{tabular}}} & Money Stolen                 & {\color[HTML]{000000} € 3444 | € 3563}    & € 3399 | € 3453                           & {\color[HTML]{009901} € 3482 | € 3511}    & {\color[HTML]{FE0000} € 2997 | € 3018}    & € 2747.48            \\ \hline
\multicolumn{1}{c|}{}                                                                                                & Injection Rate               & 89.72\% | 90.33 \%                        & {\color[HTML]{009901} 89.41\% | 90.31 \%} & 89.14\% | 90.42 \%                        & {\color[HTML]{FE0000} 82.99\% | 84.22 \%} & 87.44\%              \\
\multicolumn{1}{c|}{}                                                                                                & Time-to-Detection {[}0-10{]} & {\color[HTML]{009901} 7.99 | 8.32}        & {\color[HTML]{000000} 7.05 | 7.12}        & 7.87 | 8.01                               & {\color[HTML]{FE0000} 7.03 | 7.43}        & 7.18                 \\
\multicolumn{1}{c|}{\multirow{-3}{*}{\begin{tabular}[c]{@{}c@{}}Attacker's \\ Knowledge: \\ 320 users\end{tabular}}} & Money Stolen                 & € 3299 | € 3314                           & {\color[HTML]{009901} € 3348 | € 3466}    & € 3335 | € 3417                           & {\color[HTML]{FE0000} € 3240 | € 3333}    & € 2747.48            \\ \hline
\end{tabular}
}
\label{table:FRAUDability_results_3}
\end{table*}

\end{document}